\documentclass[prc,showpacs,preprintnumbers,unsortedaddress,amsmath,amssymb,floatfix]{revtex4}
\usepackage{graphicx}
\usepackage{bm}
\usepackage{graphicx}

\begin{document}

\title{Effects of the Tensor Force on the Multipole Response in Finite Nuclei}

\author{Li-Gang Cao$^{1,2,3,4,5}$, G. Col$\grave{\text{o}}$$^{1,2,5}$,
H. Sagawa$^{6}$, P.F. Bortignon$^{1,2}$, and L. Sciacchitano$^{1}$}

\affiliation{${}^1$Dipartimento di Fisica, Universit$\grave{\text{a}}$ 
degli Studi di Milano, via Celoria 16, Milano, Italy}

\affiliation{${}^2$Istituto Nazionale di Fisica Nucleare (INFN), Sez. di 
Milano, via Celoria 16, Milano, Italy}

\affiliation{${}^3$Institute of Modern Physics, Chinese Academy of
Science, Lanzhou 730000, P.R. China}

\affiliation{${}^4$ Center of Theoretical Nuclear Physics,
National Laboratory of Heavy Ion Accelerator of Lanzhou, Lanzhou
730000, P.R. China}

\affiliation{${}^5$ Kavli Institute for Theoretical Physics China, 
CAS, Beijing 100190, P.R. China}

\affiliation{${}^6$ Center for Mathematics and Physics, University of Aizu,
Aizu-Wakamatsu, Fukushima 965-8560, Japan}

\begin{abstract}
We present a thorough analysis of the effects
of the tensor interaction on the multipole response of magic
nuclei, using the fully self-consistent Random Phase
Approximation (RPA) model with Skyrme interactions.
We disentangle the modifications to the static mean field
induced by the tensor terms, and the specific features of  
the residual particle-hole (p-h) 
tensor 
interaction, for quadrupole (2$^+$), octupole (3$^-$), and
also magnetic dipole (1$^+$) responses. 
It is pointed out that the tensor force has a larger effect on the magnetic 
dipole states than on the natural parity states 2$^+$ and 3$^-$, 
especially at the mean field level.  
Perspectives for a better assessment 
of the tensor force parameters are eventually discussed. 
\end{abstract}


\maketitle

\section{Introduction}

The nuclear effective interactions like the zero-range Skyrme forces 
have been quite successful in many respects since they
have been introduced, mainly because of continuous efforts
to apply them to the study of different nuclear phenomena,
and to improve their form and their parameters. The Skyrme
forces contain typically 10 free parameters which are fitted
to reproduce empirical bulk properties of uniform nuclear matter, 
and masses and radii of a few magic nuclei. These forces, or 
at least several of the many existing parameter sets, describe
in a reasonable way the global features of the ground-state
properties along the nuclear chart (i.e., binding energies, radii,
deformations). In some cases one needs to input more information
in the parameter fitting: the Lyon forces \cite{Cha.98} have been for instance
determined by requiring a reproduction of the equation of state 
of neutron matter emerging from realistic calculations. 
Properties of excited states (vibrations, rotations) have been
studied using the Skyrme parameter sets, allowing a large
amount of physical insight. For a review on mean-field
calculations the reader can consult Ref. \cite{RMP}. 
Few ideas about selecting the Skyrme forces which have
reasonable overall performances can be found in \cite{bulk}.   
Last but not least, in the quest for a universal 
{\em local} Energy Density
Functional (EDF) for nuclei, the Skyrme framework is
often used as a starting point.

Of course, there are drawbacks and limitations of the 
mean-field approaches
based on effective forces. This issue should be
discussed together with the answer to the question whether
we need to generalize the existing Skyrme parametrizations.
In this spirit, since most of the Skyrme sets which have been
widely used are purely central, many groups have recently 
devoted much attention to the role played by the zero-range tensor 
terms that can be added to the standard Skyrme terms (see Refs. 
\cite{Bro.06,Dob.06,Col.07,Bri.07,Les.07,Gra.07,Zou.08,Zal.08}). 
This blooming of theoretical studies has followed the claim 
by the authors of Ref.~\cite{Ots.05}, that the tensor force is 
crucial for the understanding of the evolution of the
single-particle energies in exotic nuclei. 

Certainly, the mentioned studies have elucidated that the
tensor force does play a role. At the same time, the 
single-particle states are not the right observable on which 
to constrain an effective force or an effective functional, 
since for these states other correlations beyond mean-field
are expected to manifest themselves. 

It is certainly quite timely to analyze in detail how the
tensor terms added to a standard Skyrme force, or the new
parameter sets introduced in \cite{Les.07} (fitted with
the tensor included), behave as far as the excited states
are concerned. The study of nuclear collective vibrations,
within self-consistent Random Phase Approximation (RPA) on
top of Hartree-Fock (HF), is a well-defined framework
which has allowed testing the effective Skyrme sets for
many years. Recently, we have developed fully 
self-consistent RPA \cite{comex2}, and quasi-particle RPA (QRPA) \cite{Jun}, 
schemes. The full self-consistency, in the nonrelativistic
framework, is essential to respect theorems concerning sum rules
associated with appropriate operators. In this paper,
we extend these schemes by including the tensor terms
both in HF and in RPA as terms of the residual interaction,
and we analyze the results in the case of the multipole
response of magic nuclei. 

Our work is the first one which attempts a detailed explanation
of the effects induced by the tensor force on the 
response of finite nuclei within the Skyrme framework. 
However, the fact that the subject is of interest at present
is testified by recent papers that are devoted to 
related topics. V. De Donno {\it et al.} \cite{VDD} have studied the
low-lying magnetic excitations in several nuclei, either within
a phenomenological approach or by using the Gogny force. 
The authors of \cite{Dav.09} have investigated self-consistently, 
using the Skyrme forces and the zero-range tensor terms, the
response of uniform matter. 

The outline of our paper is the following. Since Skyrme-RPA is
a well-known framework we will survey very briefly our
formalism in Sec. \ref{theo}. More emphasis is given to the
discussion of the results, for quadrupole, octupole and
magnetic dipole states, in Sec. \ref{results}. An overall
conclusion, together with some perspectives for future work
on the tensor force and on possible constraints on its 
parameters, is given in Sec. \ref{conclu}. The main technical 
part of this work, namely the evaluation of the p-h matrix
elements of the tensor interaction, is described in Appendix 
\ref{appa}, while Appendix \ref{appb} describes and approximate
yet quite useful separable approximation for the tensor force 
that is used as a guideline to understand some features of
our numerical results.

\section{Formalism}
\label{theo}

As mentioned in the Introduction, Skyrme-RPA theory is well 
known since many years, especially in its matrix formulation.
We have been using for several applications a scheme which is fully
self-consistent, that is, where 
there is no approximation in the residual interaction since all
its terms are taken into account including the two-body
spin-orbit and Coulomb. In our scheme we use box boundary conditions. 
In other words, the continuum is discretized. In the present case 
the box radius is set at 15 fm for the Ca isotopes and at 20 fm for 
$^{208}$Pb. 

After solving the HF equations in coordinate space and determining,
together with the mean field, the unoccupied states as eigenstates
of this mean field in the mentioned large box, we build up a 
model space of p-h configurations with given multipolarity
$J^\pi$ and we write and solve the RPA matrix equation in that space.
The model space includes the configurations built up with all
hole states, and with the particle states labelled by $n_p$, $l_p$ 
and $j_p$ (number of radial nodes, spatial and total angular 
momentum); $l_p$ and $j_p$ take all possible values which are
allowed by selection rules, while $n_p$ varies between 
$n_{\rm max}+1$ ($n_{\rm max}$ is the number of nodes of the
last occupied state) and $n_{\rm max}+1+\Delta n$. 

The value of $\Delta n$ is chosen large enough, so that 
the isoscalar (IS) EWSR for the multipoles 2$^+$, 3$^-$ exhaust 
practically all ($\sim$ 97-99\%) the double commutator (DC) value.
In particular, we use $\Delta n$=10 unoccupied shells
in the case of the 2$^+$ and 
1$^+$ calculations in $^{40}$Ca and $^{48}$Ca, 
$\Delta n$=9 unoccupied shells in the case of the 3$^-$ calculations
in the same nuclei, and $\Delta n$=8 unoccupied shells
for $^{208}$Pb.

The eigenvalues $E_n$ and eigenvectors 
$\vert n \rangle$ of the RPA equations allow calculating
the response function to different operators. In particular we 
are going in what follows to show figures with the strength 
function associated with the operator $\hat F$, namely
\begin{equation}
S(E) = \sum_n |\langle n \vert \hat F \vert 0 \rangle|^2 
\delta(E-E_n), 
\end{equation}
where the sum spans the whole RPA spectrum. 
In the case of discretized RPA one has in reality sharp peaks but in the
figures we shall display $S(E)$ smoothed out using Lorentzian
functions (having 1 MeV width). We shall use isoscalar quadrupole
and octupole operators, that is, 
\begin{equation}
\hat F_L = \sum_{i=1}^A \sqrt{2L+1}\ r^L_i Y_{LM} (\hat r_i)
\label{natpar_op}
\end{equation}
with $L$=2 and 3, respectively. We focus our study on the 1$^+$ spin-flip
states as well. In this case the following operators are used,
namely
\begin{equation}
\hat F_J (IS) = \sum_{i=1}^{A}\frac{g^{IS}
e\hbar}{2mc}\sqrt{J(2J+1)}\ \sigma_{i}^{\mu}Y_{00}(\widehat{r}_i)
\label{m1is}
\end{equation}
in the isoscalar case, and 
\begin{equation}
\hat F_J (IV) = \sum_{i=1}^{A}\frac{g^{IV}
e\hbar}{2mc}\sqrt{J(2J+1)}\ \sigma_{i}^{\mu}Y_{00}(\widehat{r}_i)\tau_{i}^{z}
\label{m1iv}
\end{equation}
in the isovector case. We shall consider the case $J$=1. In the above 
equations the nuclear magneton
$\mu_N=\frac{e\hbar}{2mc}$ appears together with the quantities 
$g^{IS}=\frac{1}{4}(g_s^n+g_s^p-1)$=0.19, and $g^{IV}=\frac{1}{4}
(g_s^n-g_s^p+1)$=-2.10.

We note that the tensor contribution must not alter the 
DC value for the EWSRs in the case of the operators (\ref{natpar_op}), 
for all values of $L$. It does change, instead, the value of the EWSRs
in the case of the operators (\ref{m1is}) and (\ref{m1iv}) (cf., e.g., 
Ref. \cite{Liu.91}). 

Two sets of Skyrme parameters are employed in the calculations
presented below. On the one hand, we have chosen the set SLy5 \cite{Cha.98}
and supplemented it with tensor terms characterized by the values
chosen in \cite{Col.07} and used in \cite{Zou.08} as well. The analysis
of these results will elucidate qualitatively the role played by the
tensor terms, but in order to compare with the case of a force having
all parameters fitted on equal footing (including those of the tensor
terms) we have performed also calculation with the set T44 introduced
in Ref. \cite{Les.07}. For the reader's convenience, all the parameters
are shown in Table \ref{table_param}.   

\section{Results}
\label{results}

\subsection{Quadrupole response}
\label{quadrupole}

\begin{figure}[hbt]
\includegraphics[width=0.495\textwidth]{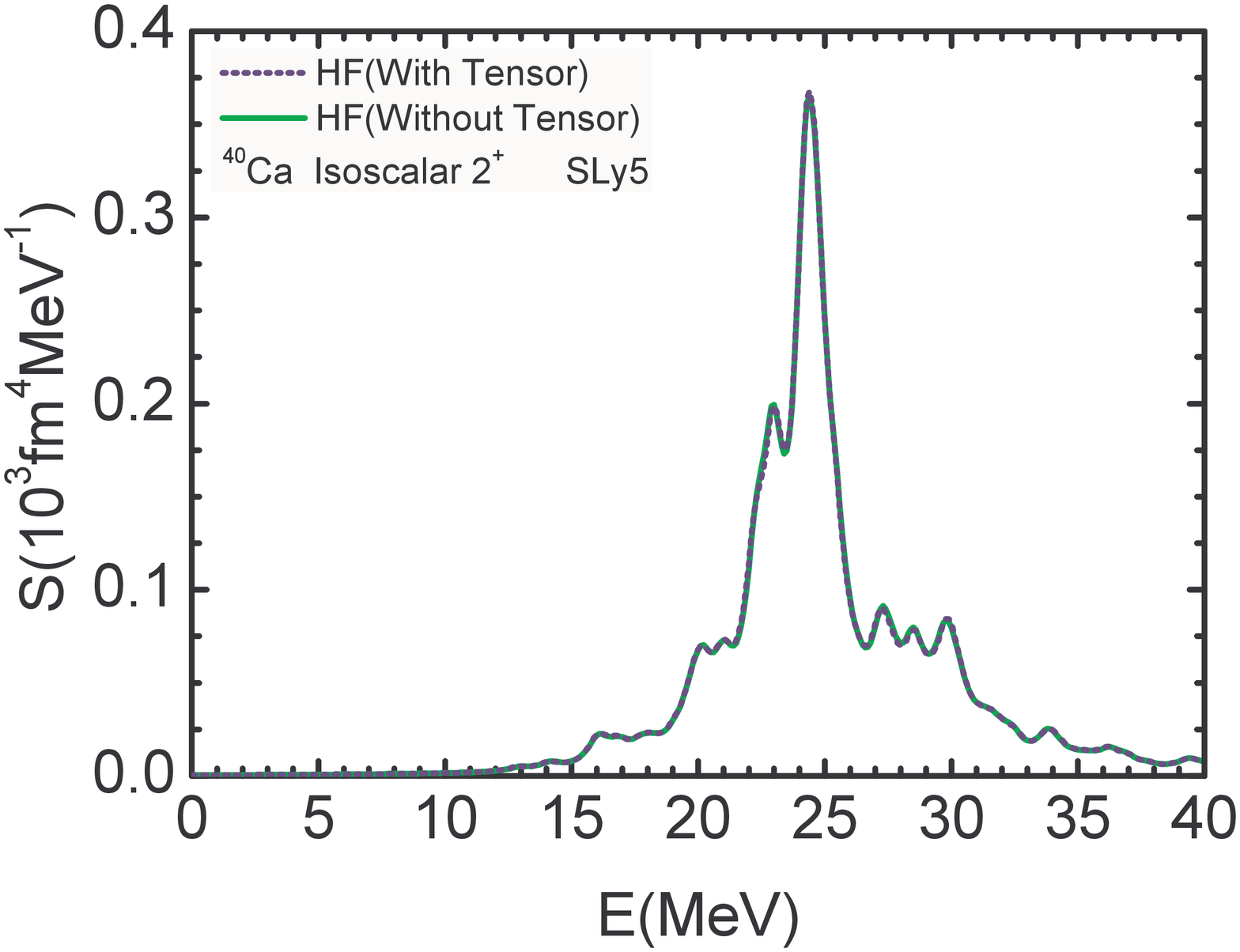}
\includegraphics[width=0.495\textwidth]{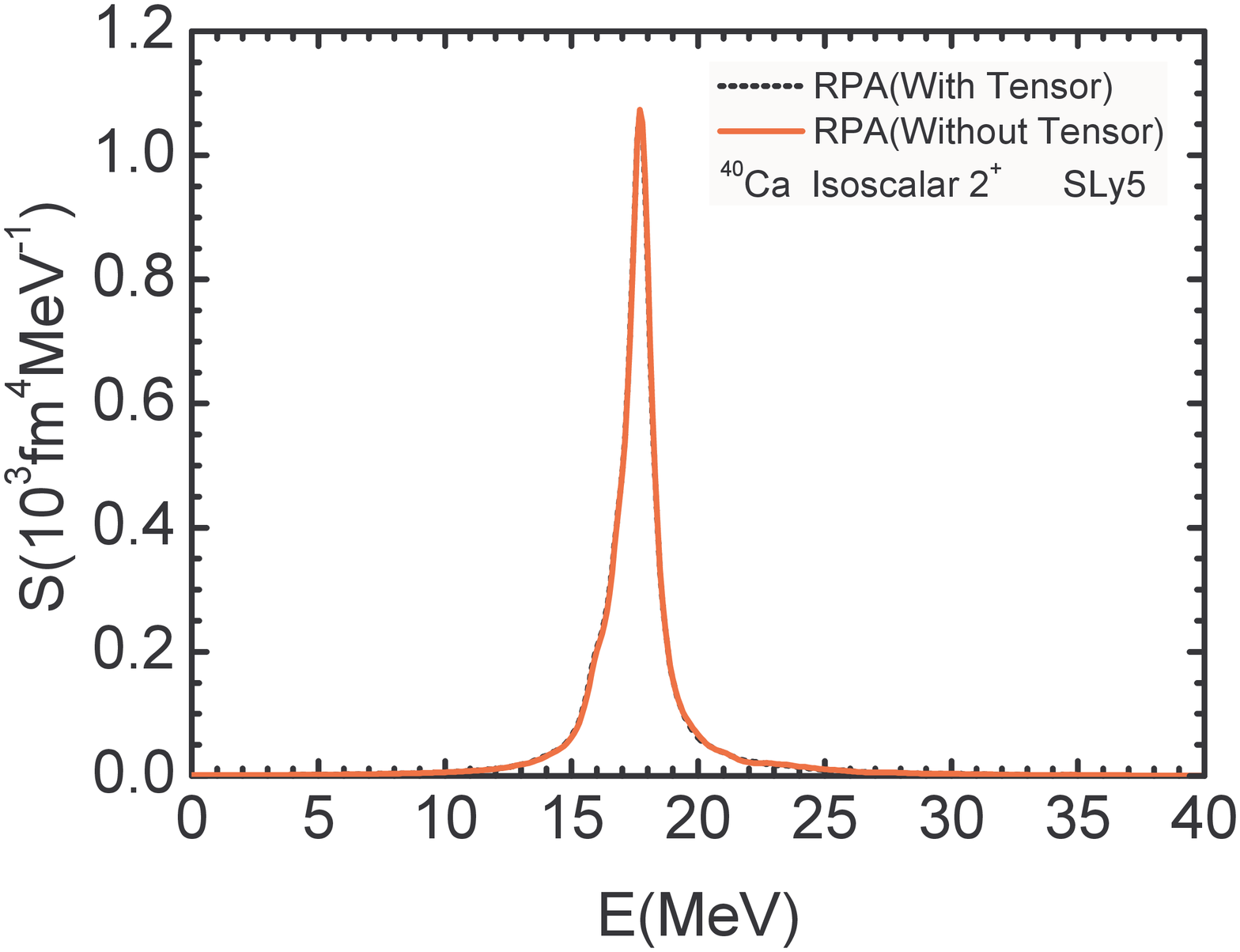}
\vglue -2.cm
\includegraphics[width=0.495\textwidth]{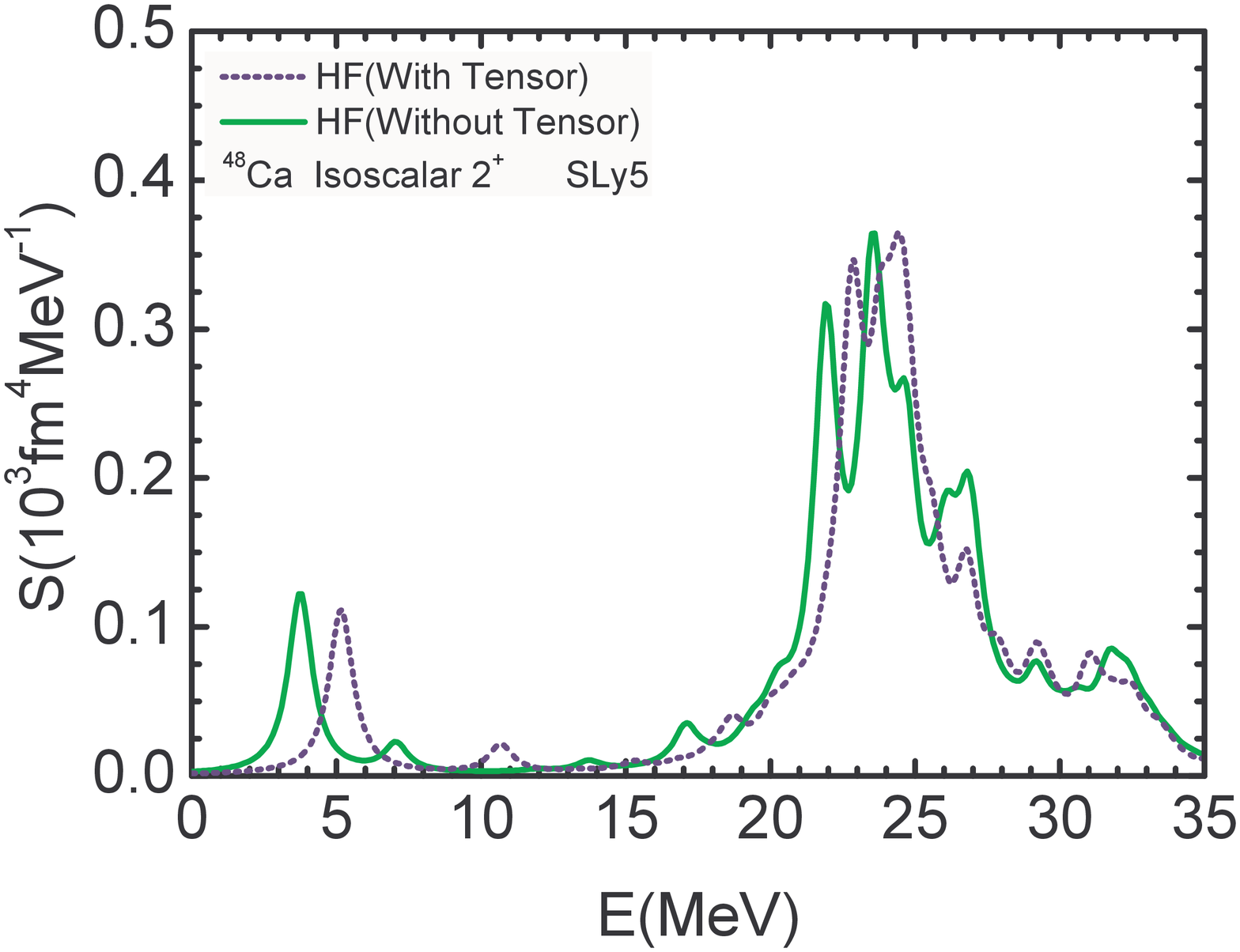}
\includegraphics[width=0.495\textwidth]{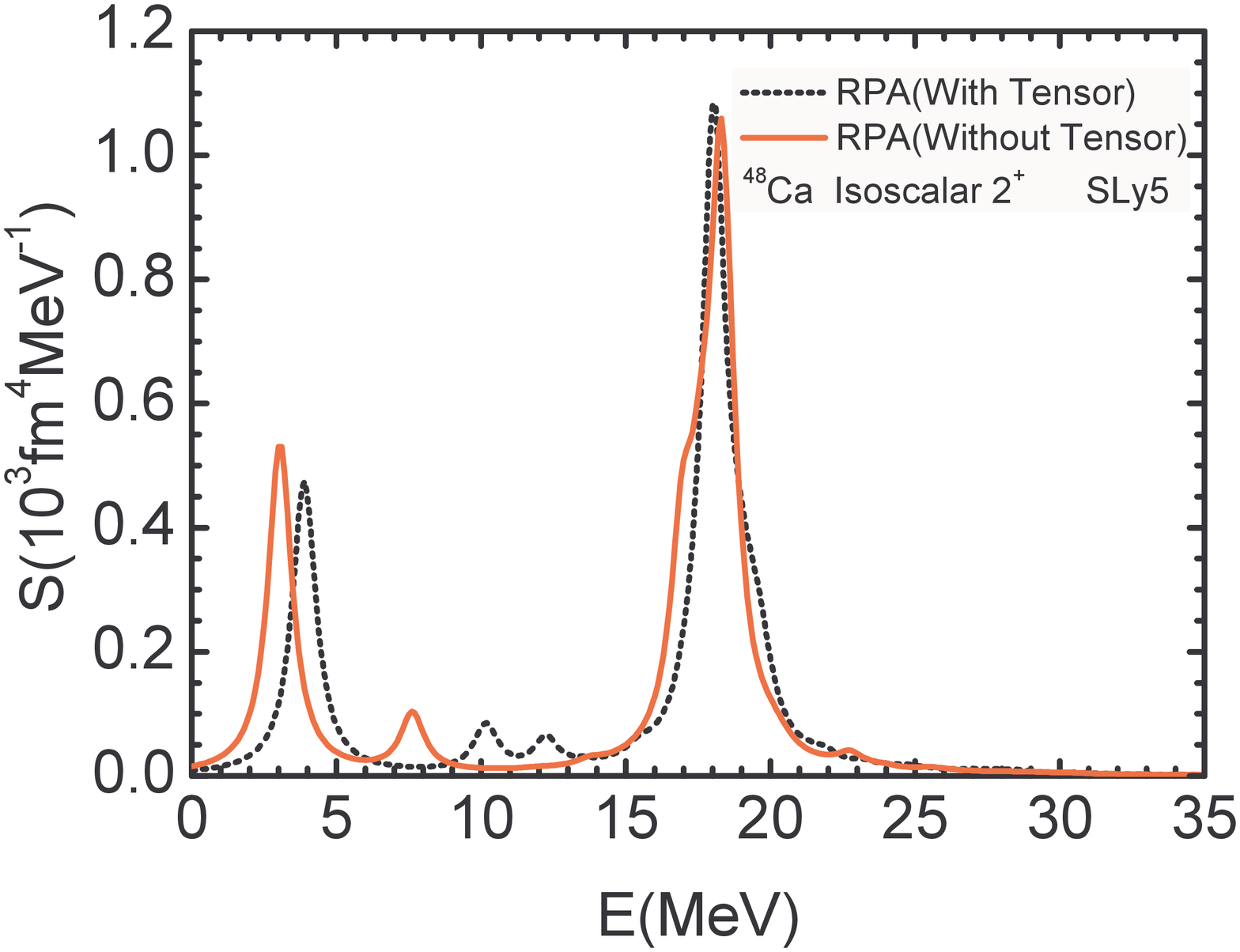}
\vglue -2.cm
\includegraphics[width=0.495\textwidth]{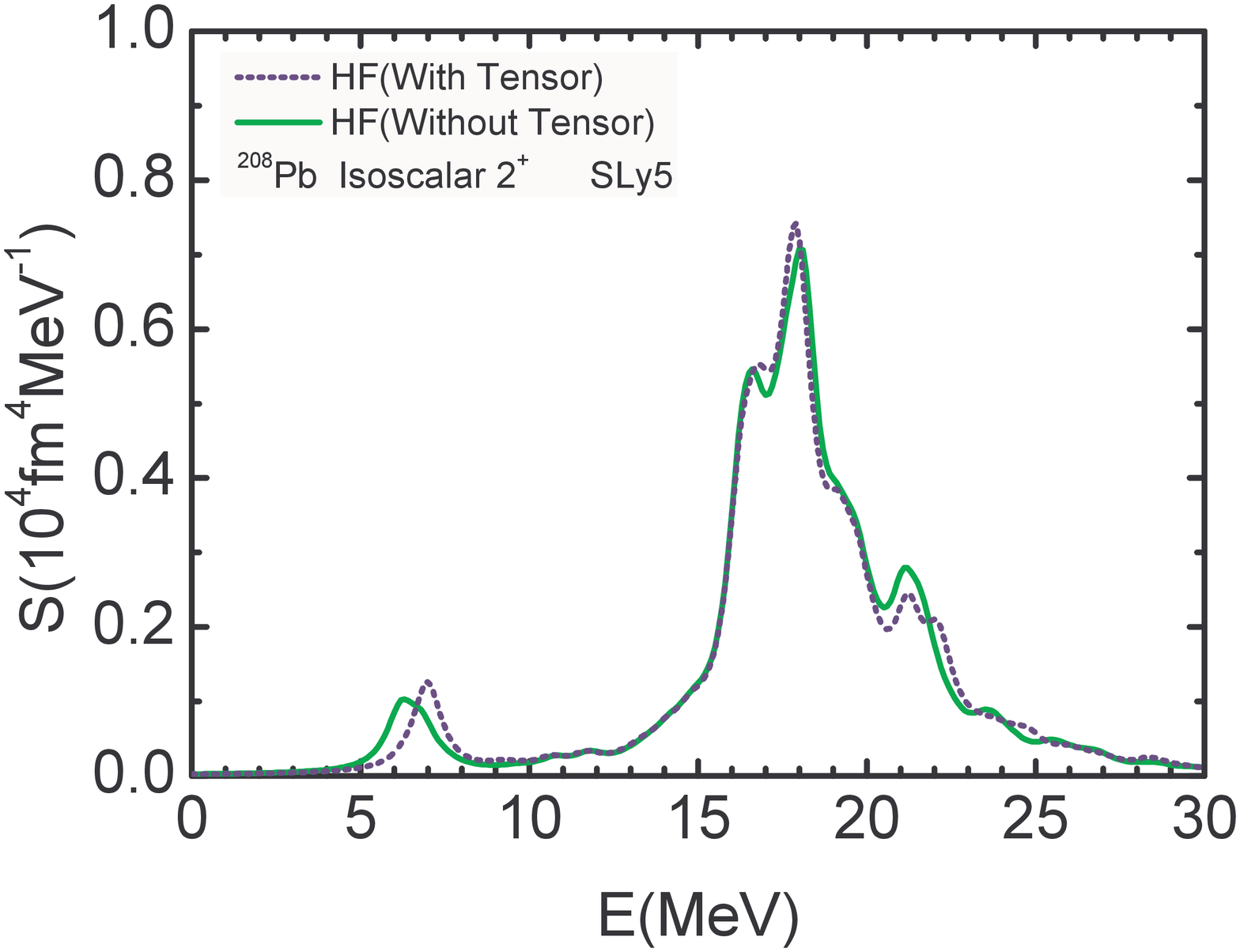}
\includegraphics[width=0.495\textwidth]{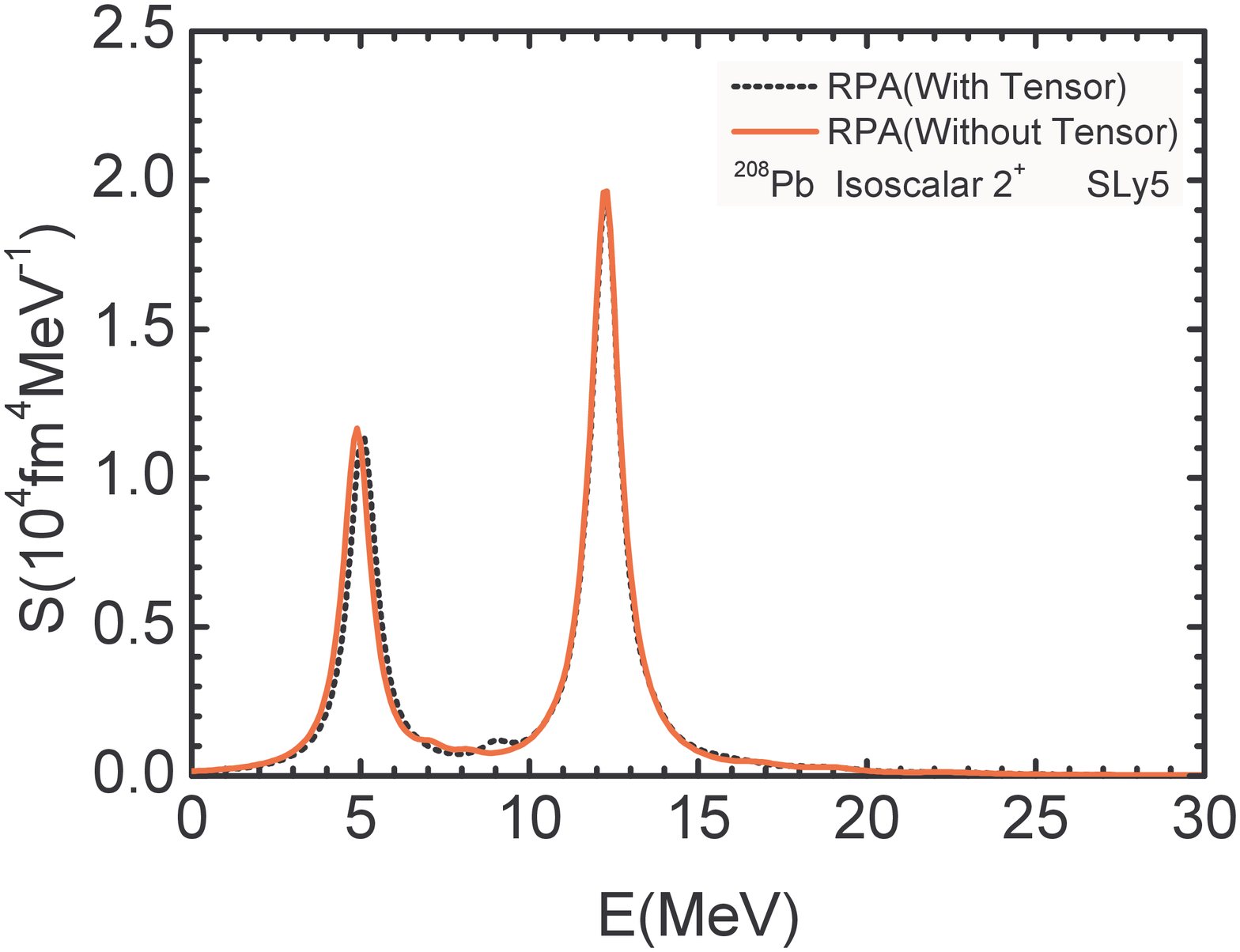}
\vglue -2.cm
\caption{
Unperturbed (left panels) and RPA (right panels) strength functions  
associated with the isoscalar quadrupole operator, in the case of 
$^{40}$Ca, $^{48}$Ca and $^{208}$Pb. We display results both with and
without the inclusion of the tensor force in the case of the
force SLy5 plus the $T$ and $U$ parameters of \cite{Col.07} 
(first column of Table \ref{table_param}). The discrete RPA
peaks have been smeared out by using Lorentzian functions which
have 1 MeV width.} \label{Fig.1}
\end{figure}

Fig. \ref{Fig.1} displays in the upper panel the IS response of the 
nucleus $^{40}$Ca. For a $\vec l\cdot\vec s$-closed nucleus (that is,
a nucleus where both spin-orbit partner states are occupied), and a 
natural-parity excitation spectrum, we do not expect that the
tensor interaction plays any significant role. 
In particular, for the ground state, the changes in the spin-orbit
splittings due to the tensor terms have been written in previous works 
and they depend on the so-called spin-orbit densities defined as 
\begin{equation}\label{Jq}
J_q(r)=\frac{1}{4\pi r^3}\sum_i (2j_i+1)[j_i(j_i+1)-l_i(l_i+1)-\frac{3}{4}]
R_i^2(r),
\end{equation}
where the isospin quantum number $q=0(1)$ labels neutrons (protons) and
the index $i=n,l,j$ runs over all states having the given $q$. 
$R_i(r)\equiv\frac{u_i(r)}{r}$ is the radial part of the wavefunction. 
So, our expectation is obviously based, in the unperturbed case, on 
the fact that $J_q$ is negligibly small (both for neutrons and for 
protons) in the case at hand. In fact, the 
microscopic calculations show that the results are exactly the same, 
with and without the tensor force. 

For comparison, we have computed the same IS response in 
the nucleus $^{48}$Ca and the results are shown in the middle
panel of Fig. \ref{Fig.1}. The response in the giant resonance region is not
affected strongly by the inclusion of the tensor force. Although 
some changes are found in the high-energy part of the strength 
function, the main impact of the tensor terms is visible in the 
states below the isoscalar giant quadrupole resonance (ISGQR). 
For the sake of clarity, the properties of the lowest 2$^+$  
states are shown in Table \ref{table_2plus}. We have analyzed 
the changes of these properties induced by the tensor 
force. With the interaction SLy5, and the tensor
parameters $T$ and $U$ already used in Refs. \cite{Col.07,Zou.08}, the 
spin-orbit splittings are increased. Normally, at least for
well-bound states, the splittings should increase more for larger
values of $l$. Consequently, the low-lying
2$^+$ state in $^{48}$Ca, which is mainly due to the neutron f$_{7/2}\rightarrow$
p$_{3/2}$ transition, is pushed upward in the unperturbed response
by the tensor force. We denote this shift by $\Delta E_{\rm HF}$. 
The effect of the residual tensor force $V_{\rm tensor}$ included
in RPA can be estimated by means of
\begin{equation}
\Delta E_{\rm RPA} \approx \Delta E_{\rm HF} + \langle V_{\rm tensor}
\rangle,
\label{DeRPA}
\end{equation}
where $\Delta E_{\rm RPA}$ indicates the difference between the RPA result
with and without the tensor force, and $\langle \rangle$ means that
we extract here an average value of the residual force.  In our 
SLy5 calculation, the value
of $\Delta E_{\rm HF}$ is 1.41 MeV
and from the shift of the RPA peak (0.83 MeV) we extract
$\langle V_{\rm tensor} \rangle$=-0.58 MeV. 

The fact that the residual force is attractive, albeit not large, 
can be understood using the argument developed in Appendix 
\ref{appb}. Using a separable approximation for the tensor
interaction, it is shown that both the tensor-even and tensor-odd 
terms produce indeed attraction in the present case, where
$T$ is positive and $U$ is negative [cf. Eqs. (\ref{appb:eq9})
and (\ref{appb:eq13})]. 

Our understanding needs to be confirmed by another example. In the
lower panel of Fig. \ref{Fig.1}, the IS response for
the nucleus $^{208}$Pb is displayed. There is essentially 
no effect of the tensor force in the giant resonance region. 
The low-lying 2$^+$ state is moved upward in the calculation with tensor by 
$\Delta E_{\rm RPA}$=0.17 MeV. However, if we look at the 
centroid of the low-energy unperturbed response, the shift due
to the tensor force is $\Delta E_{\rm HF}$=0.60 MeV. 
Then, from Eq. (\ref{DeRPA}) we extract a value of 
$\langle V_{\rm tensor} \rangle$=-0.43 MeV, that is, consistent
with what extracted in the case of $^{48}$Ca and with the
qualitative argument of Appendix \ref{appb}. 

The inclusion of the tensor force on top of SLy5, with the
parameters used in our previous works  \cite{Col.07,Zou.08}, 
does not improve the agreement with experimental values. 
It is desirable to check the behavior of a different 
interaction, and we have chosen one of those which have been
fitted including the tensor terms in Ref. \cite{Les.07},
that is, the set named T44. 

In this case, not so much emphasis should be put on the 
discussion of the effect of the tensor terms separately 
from that of 
other Skyrme parameters, since all parameters have been
fitted together. We just briefly point out that the
tensor residual interaction is almost negligible in
this case. In fact, for the low-lying 2$^+$ state of 
$^{48}$Ca calculated with T44, the values of  
the quantities defined in Eq. (\ref{DeRPA}) are 
$\Delta E_{\rm HF}$=-0.13 MeV, $\Delta E_{\rm RPA}$=-0.18 
MeV and $\langle V_{\rm tensor} \rangle$=-0.05 MeV. 
The same values are $\Delta E_{\rm HF}$=-0.54 MeV, 
$\Delta E_{\rm RPA}$=-0.55 MeV and $\langle V_{\rm tensor} 
\rangle$=-0.01 MeV in the case of $^{208}$Pb. We should
also mention that in the case of infinite
matter \cite{Dav.09} the effect of the tensor terms has been
found to be small, in the case of the T$ij$ parameter sets 
of Ref. \cite{Les.07}, 
looking at the $S=0$ response both with and without the tensor. 

In conclusion, the behavior of the tensor interaction in
the quadrupole response can be fairly well understood. Its 
effects are visible only for the low-lying
states (not for giant resonances) and are mainly related to the 
mean field since the residual tensor interaction matrix elements 
are either tiny or anyway not very large. We end this Section by adding
that the isovector response has also been computed, although it
is not displayed in our figures. We have in fact observed that
the centroid energy of the (high-lying) isovector 
strength may change by few hundreds of
keV and only in few cases by about $\approx$ 1 MeV when
the tensor force is included.

\subsection{Octupole response}
\label{octupole}

\begin{figure}[hbt]
\includegraphics[width=0.495\textwidth]{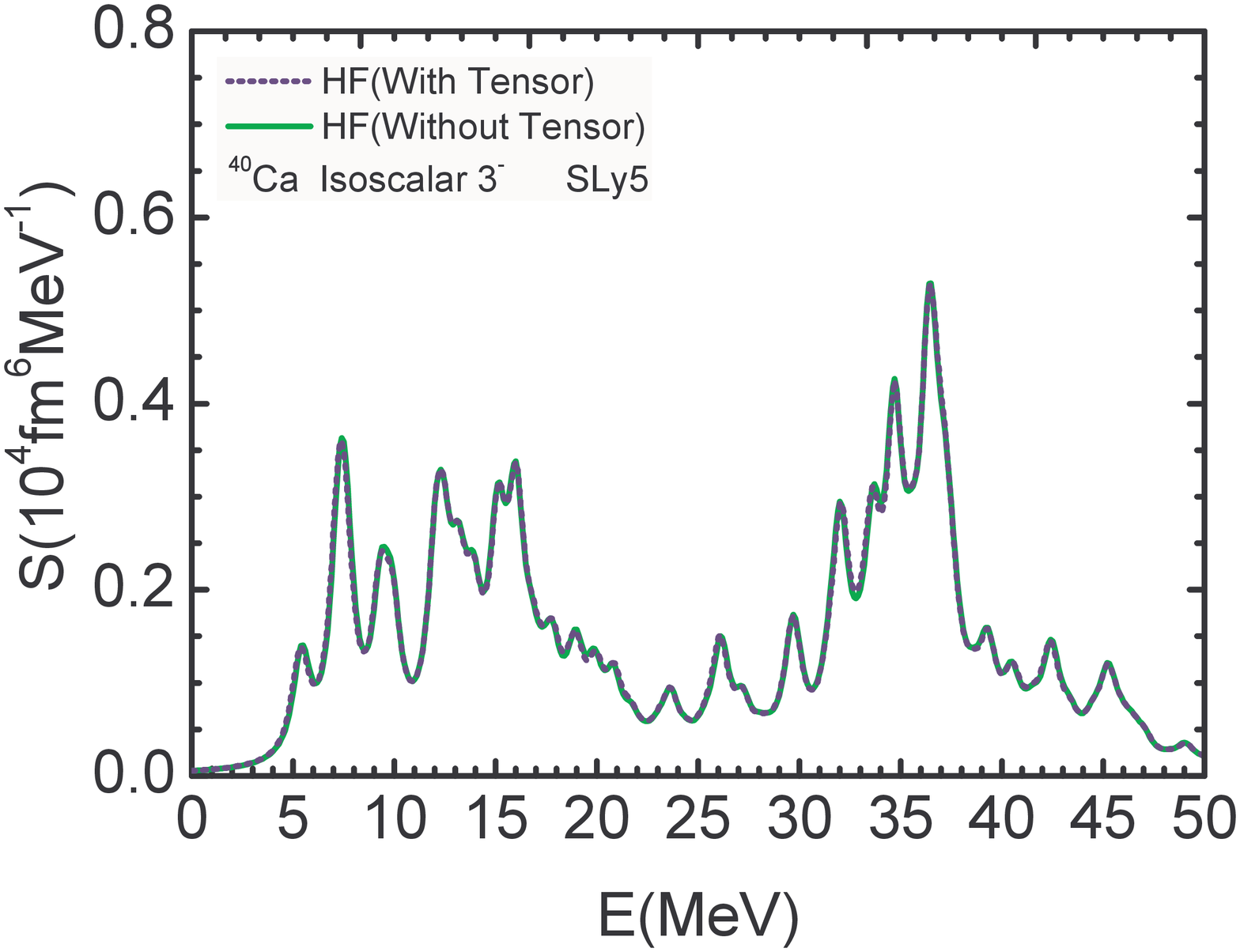}
\includegraphics[width=0.495\textwidth]{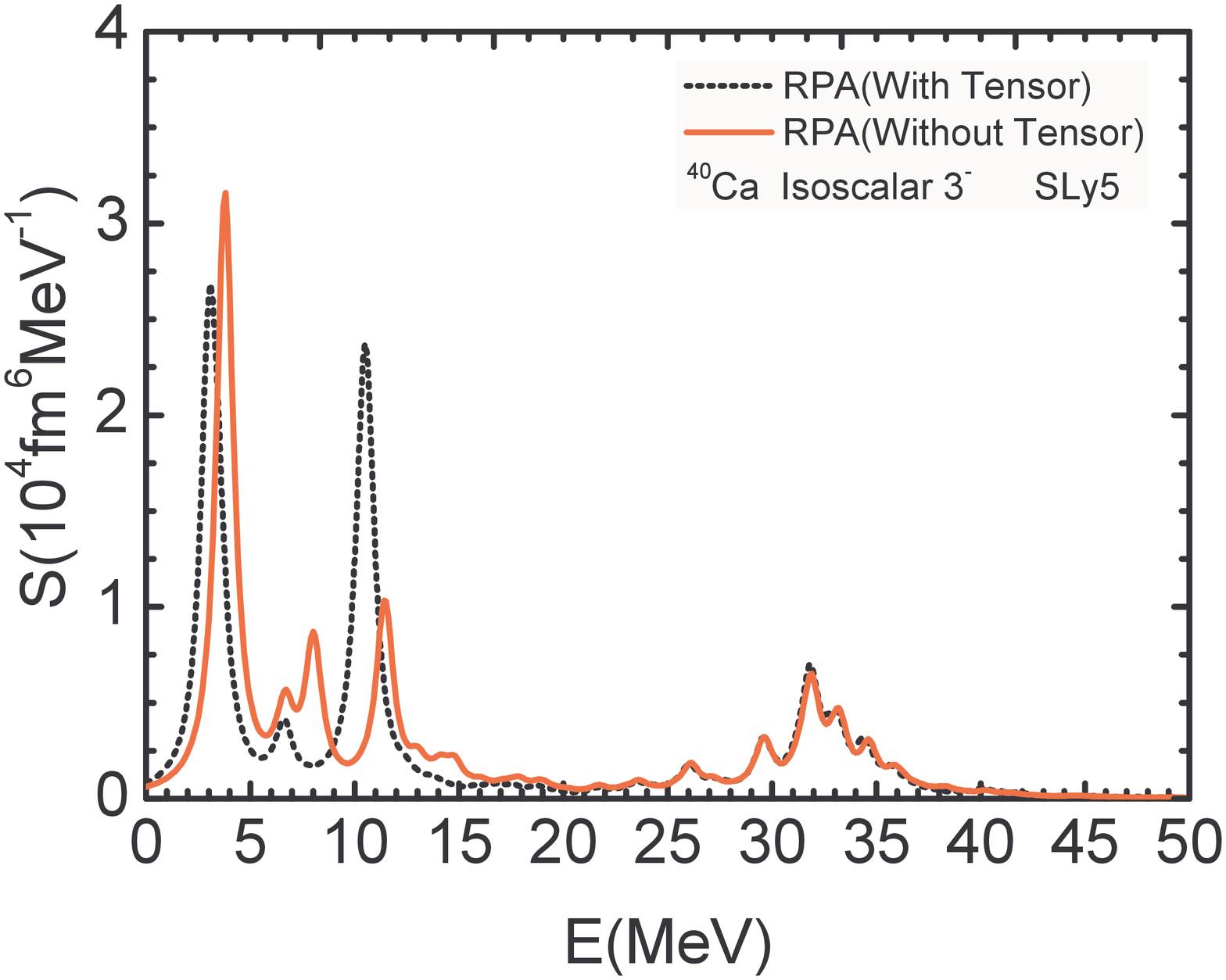}
\vglue -2.0cm
\includegraphics[width=0.495\textwidth]{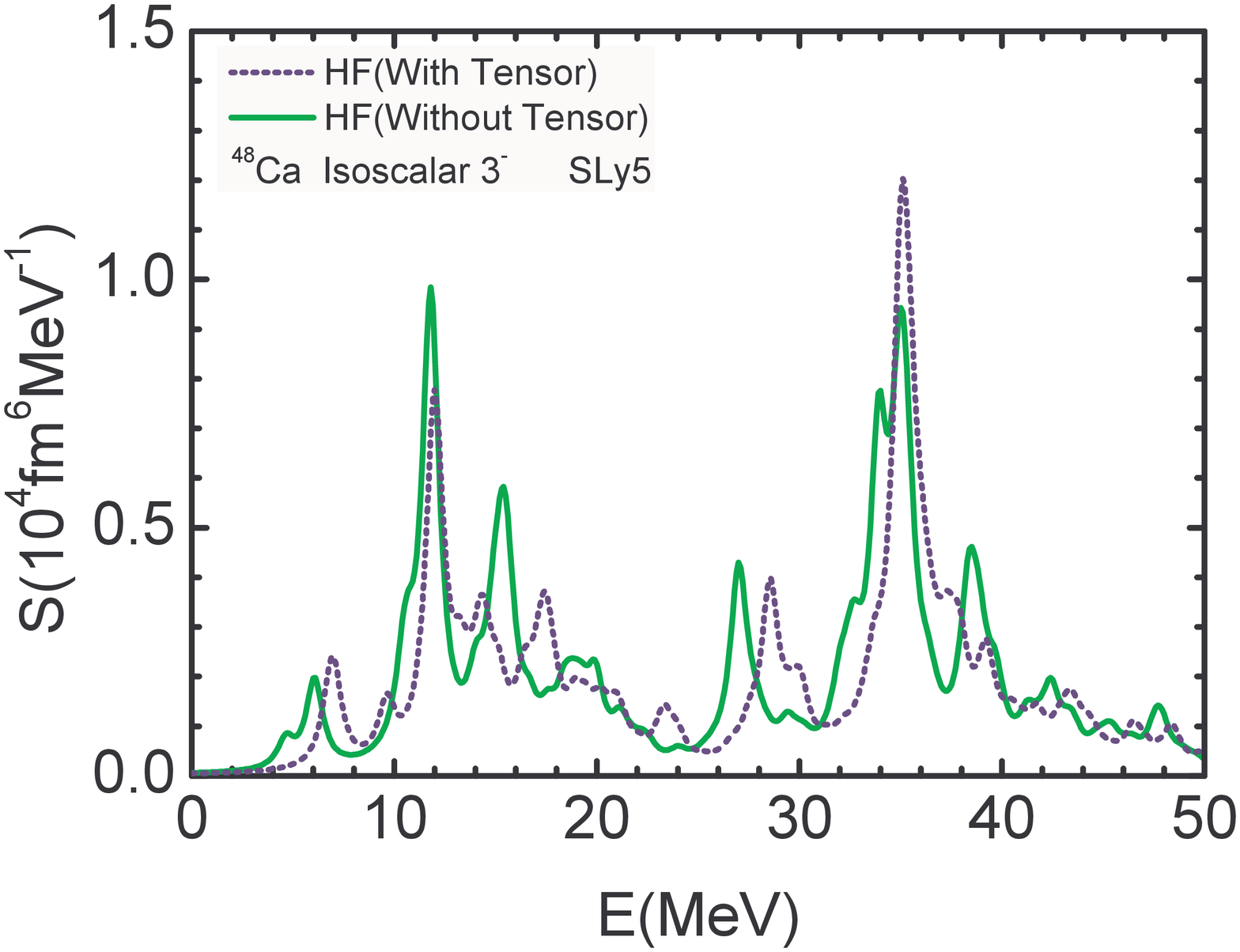}
\includegraphics[width=0.495\textwidth]{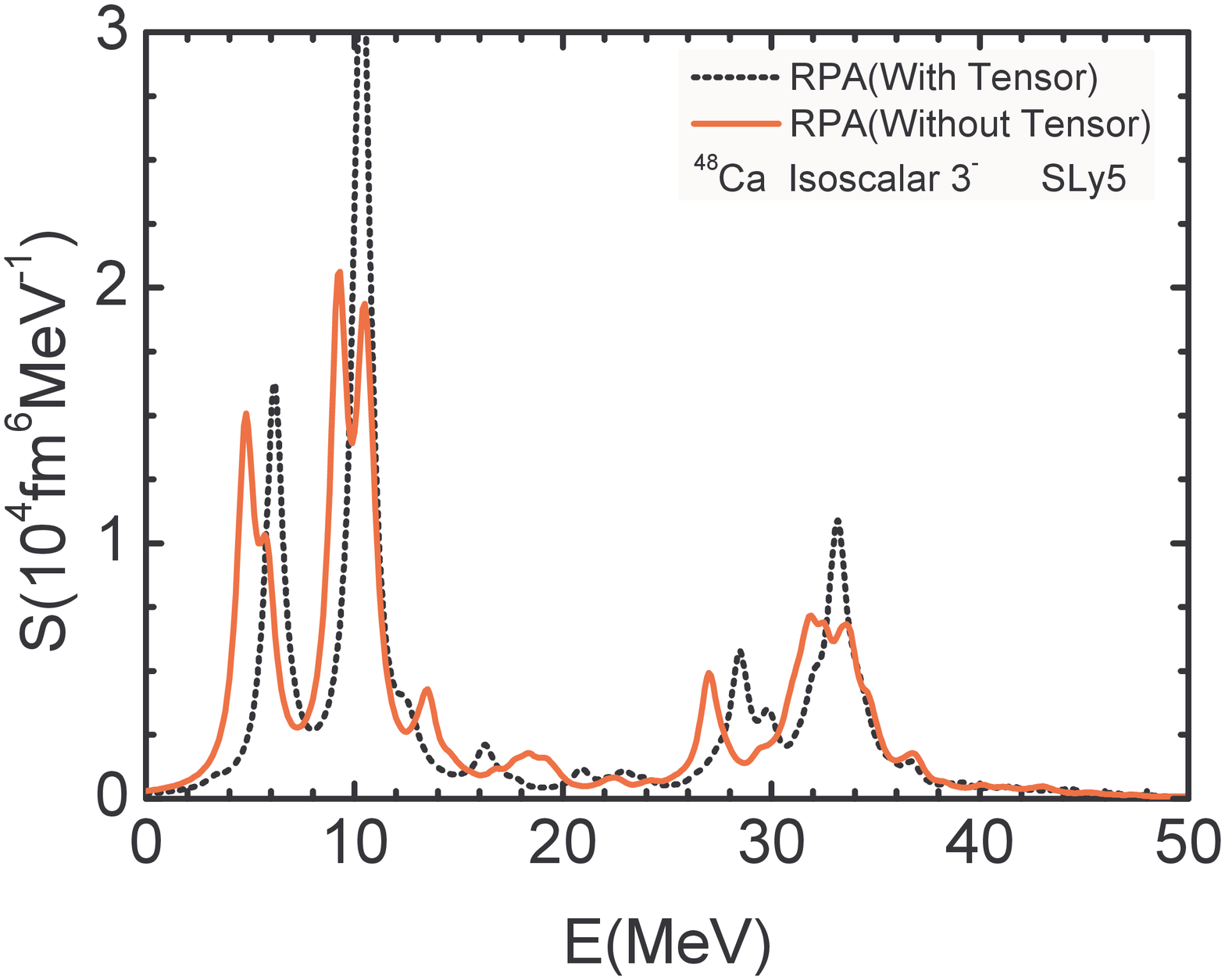}
\vglue -2.0cm
\includegraphics[width=0.495\textwidth]{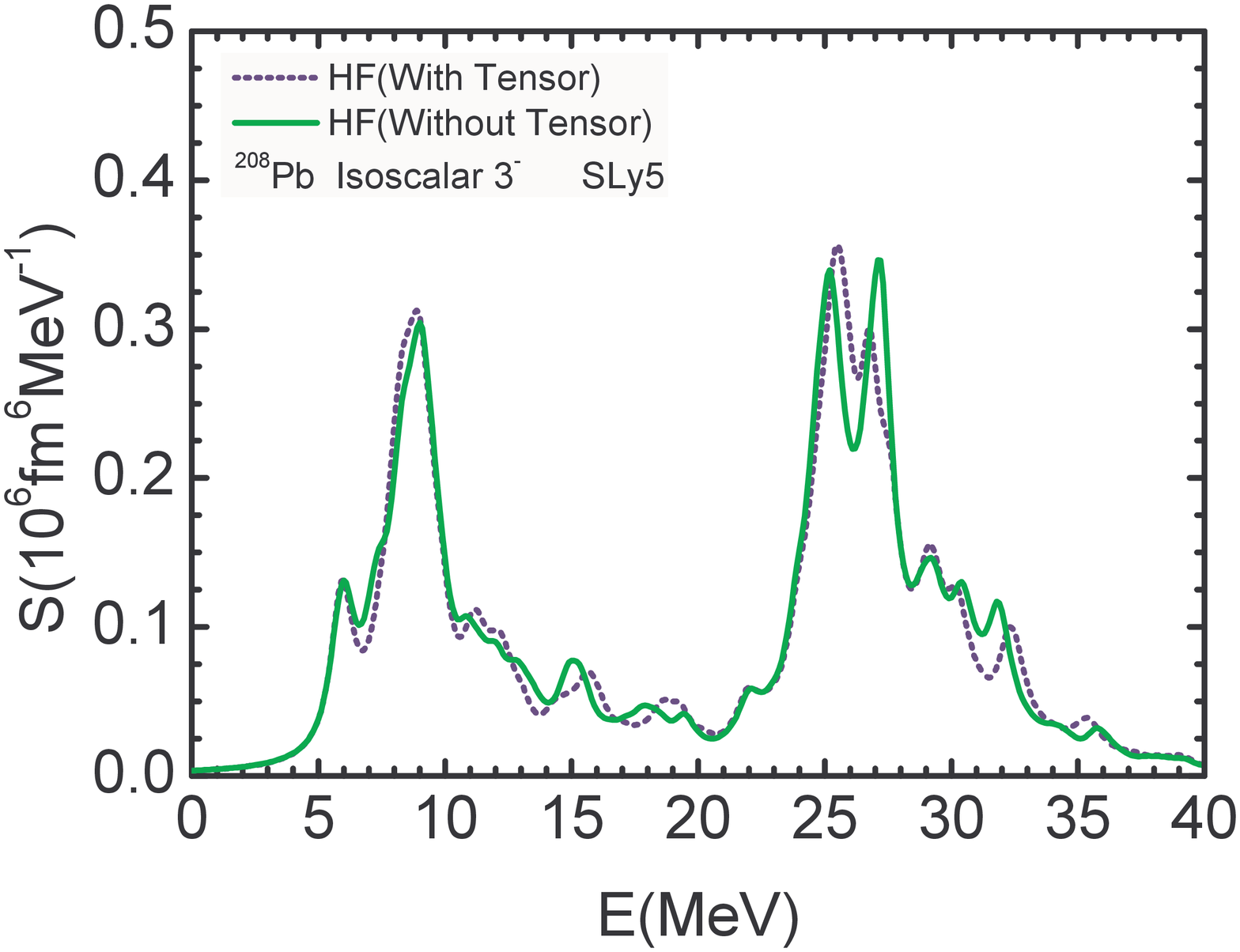}
\includegraphics[width=0.495\textwidth]{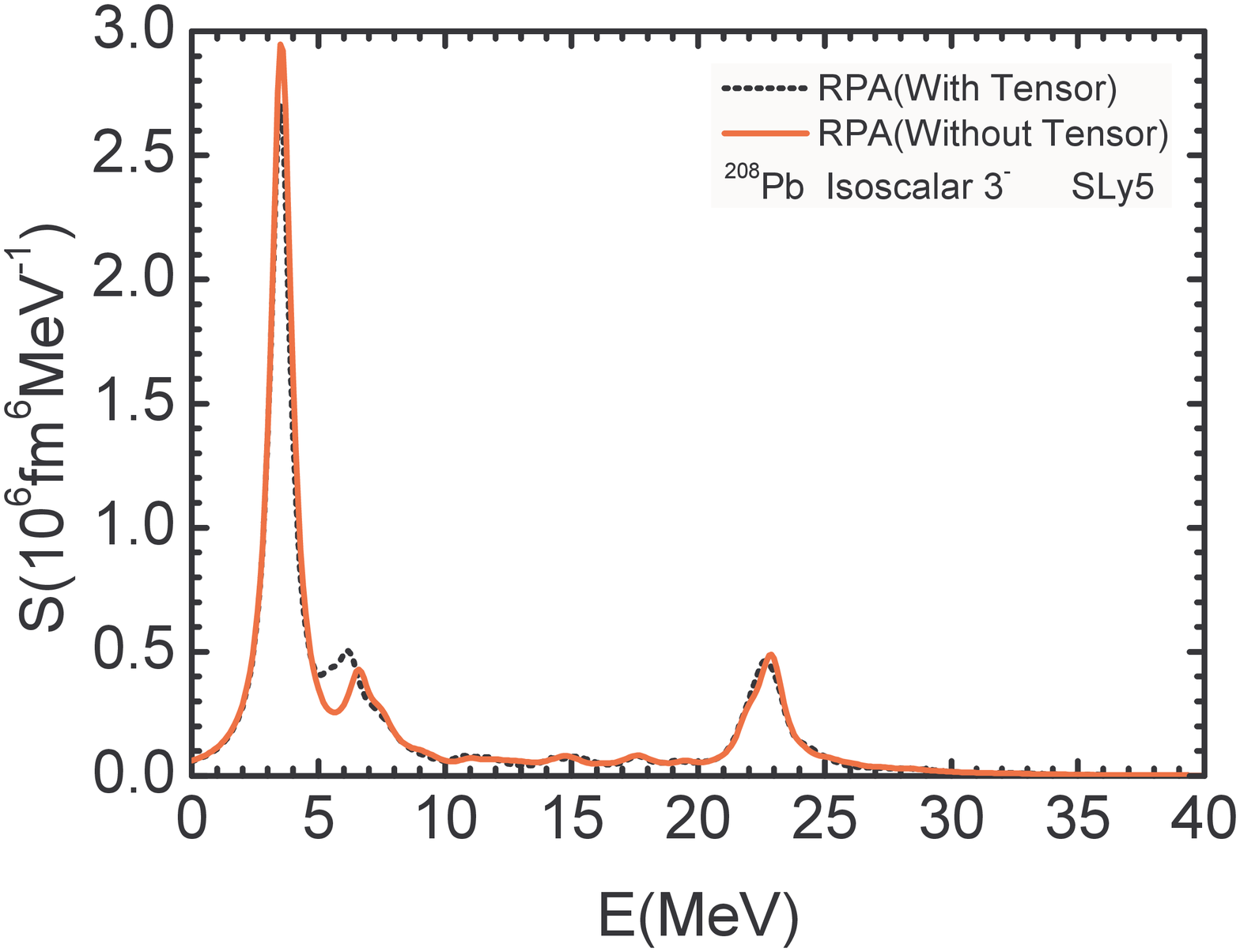}
\vglue -2.0cm \caption{The same as Fig. \ref{Fig.1} in the case of
the octupole response.} \label{Fig.3}
\end{figure}

In a similar style as for the quadrupole case, the IS octupole responses
of the nuclei $^{40}$Ca, $^{48}$Ca and $^{208}$Pb are displayed in
the upper, middle and lower panels of Fig. \ref{Fig.3}, respectively. Also
in the present case, we do not discuss the IV strength (for
which a comment similar to the one made at the end of the previous
subsection applies) but only the IS strength. 

The detailed behavior of the low-lying states is shown in Table 
\ref{table_3minus}. In the case of $^{40}$Ca, as we have already commented
in the previous subsection, there is no effect of the tensor terms at
the mean field level since $J_p$ and $J_n$ are negligibly small. In the case
of SLy5 plus the $T$ and $U$ values of \cite{Col.07,Zou.08}, the value
of $\langle V_{\rm tensor} \rangle$ can be therefore deduced from the
shift of the RPA energy of the lowest 3$^-_1$ state, 
and it is -0.76 MeV, not much different from
the values obtained in the case of the low-lying 2$^+$ (cf. the previous
subsection).  

In the case of $^{48}$Ca, there is no systematic behavior in the shifts
of the unperturbed p-h transitions when the tensor terms are included.
In this nucleus $J_p$ is negligibly small while $J_n$ is positive due
to the neutron f$_{7/2}$ contribution. With the force SLy5, and 
the parameters $T$ and $U$ of \cite{Col.07,Zou.08}, 
the proton spin-orbit splittings are reduced, whereas the neutron ones are
enlarged. The 3$^-_1$ state, which appears without tensor at 4.78 MeV 
as it is shown in Table \ref{table_3minus}, 
has the proton s$_{1/2}\rightarrow$ f$_{7/2}$ and d$_{3/2}\rightarrow$
f$_{7/2}$ as main components, and both are pushed upward in energy when
the tensor is included. This explains the large positive shift of this
RPA state, despite the (not strong) attractive contribution provided
by the residual
interaction. The same p-h configurations are also the main ones which
contribute to the second RPA state at 5.80 MeV which is visible in 
Fig. \ref{Fig.3} (in the curve corresponding to the calculation without
tensor). However, in the case of other RPA states the situation is different.
The third peak visible in Fig. \ref{Fig.3} (once more, in the curve corresponding to 
the RPA calculation without tensor) lies at 9.23 MeV, and it receives 
contribution both from
configurations which are pushed upward by the tensor inclusion, 
like the proton d$_{3/2}\rightarrow$ p$_{3/2}$ and the 
neutron f$_{7/2}\rightarrow$ g$_{9/2}$, and from
others which are pushed downward like the neutron d$_{3/2}\rightarrow$ p$_{3/2}$.
So, the inclusion of the tensor cannot produce an effect which
is understandable in simple terms. On top of this, we have verified
that the RPA states calculated with the tensor included, contain 
somewhat different admixtures of p-h configurations with respect to
those calculated without the tensor force. 

Also in $^{208}$Pb several configurations contribute to the collective 
low-lying 3$^-_1$ state. This collectivity is enough, so that the effects 
coming from unperturbed configuration which are pushed upward or downward by
the tensor terms, together with the effect of the tensor residual interaction, 
essentially cancel one another. Both from Fig. \ref{Fig.3} and 
Table \ref{table_3minus} one can see that the energies and B(E3) values in  
$^{208}$Pb are not affected significantly when tensor is included.
The conclusion which can be reached, looking at the results obtained
by using the SLy5 interaction and the tensor parameters from 
\cite{Col.07,Zou.08}, is that the effects of the tensor force on
the octupole states are rather dependent on
the specific nucleus, that is, on the relevant shell model states. 
In any case, we do not observe huge effects. 

If we use the interaction T44, these qualitative conclusions remain
valid, although the behavior of the tensor force is different in detail 
because of the contributions with different signs of $T$ and $U$, and 
their smaller absolute values. Actually, as in the quadrupole case,  
these differences do not manifest dramatically in the final results
of the low-lying states in $^{48}$Ca and $^{208}$Pb. In the case
of $^{40}$Ca, the gap in this nucleus turns out to be very small
(about 3.3 MeV) as compared to the case of standard Skyrme forces
which are fitted without tensor terms. In fact, this gap is,
for example, 
5.51 MeV with SLy5. This explains the small energy of the low-lying
3$^-$ in $^{40}$Ca. In some cases this behavior of T44 might lead to 
RPA instabilities, which, however, we have not observed by studying these
low multipole responses in finite nuclei. 

\subsection{Magnetic dipole response}
\label{dipole}

On quite general grounds, it can be expected that the effects of the 
tensor force are larger for spin (or for spin-isospin) states, both at the
mean field level since unperturbed configurations are sensitive to
the spin-orbit splittings, and also as far as RPA correlations are
concerned. Some of the authors of the present work have already 
analyzed the role of the tensor correlations in the case of charge-exchange, 
spin-isospin states like the Gamow-Teller (GT) and spin-quadrupole 
resonances \cite{Bai.09}. Tensor correlations are quite strong in 
that case, and lead to a lowering of the main GT peak by about 2 MeV
in $^{90}$Zr and $^{208}$Pb, which is accompanied by a strong repulsive
shift of a sizeable fraction of strength (pushed at energies 
higher than the giant resonance region, at 30 MeV and above). These effects are
certainly more prominent than what has been discussed in the
previous subsection for the non spin-flip transitions. Many earlier
works in the literature have emphasized the role of the tensor force
in the spin and spin-isospin channel, but not in a self-consistent
framework like in the present case or in the case of Ref. \cite{Bai.09}. 
The self-consistent calculation of the uniform matter response 
performed using the Skyrme force T44 in \cite{Dav.09}, confirms that 
larger and non-trivial effects from the tensor terms may be
expected in the spin channel. With this background, we have
made calculations for the magnetic dipole response in 
$^{48}$Ca and $^{208}$Pb. 

\begin{figure}[hbt]
\includegraphics[width=0.495\textwidth]{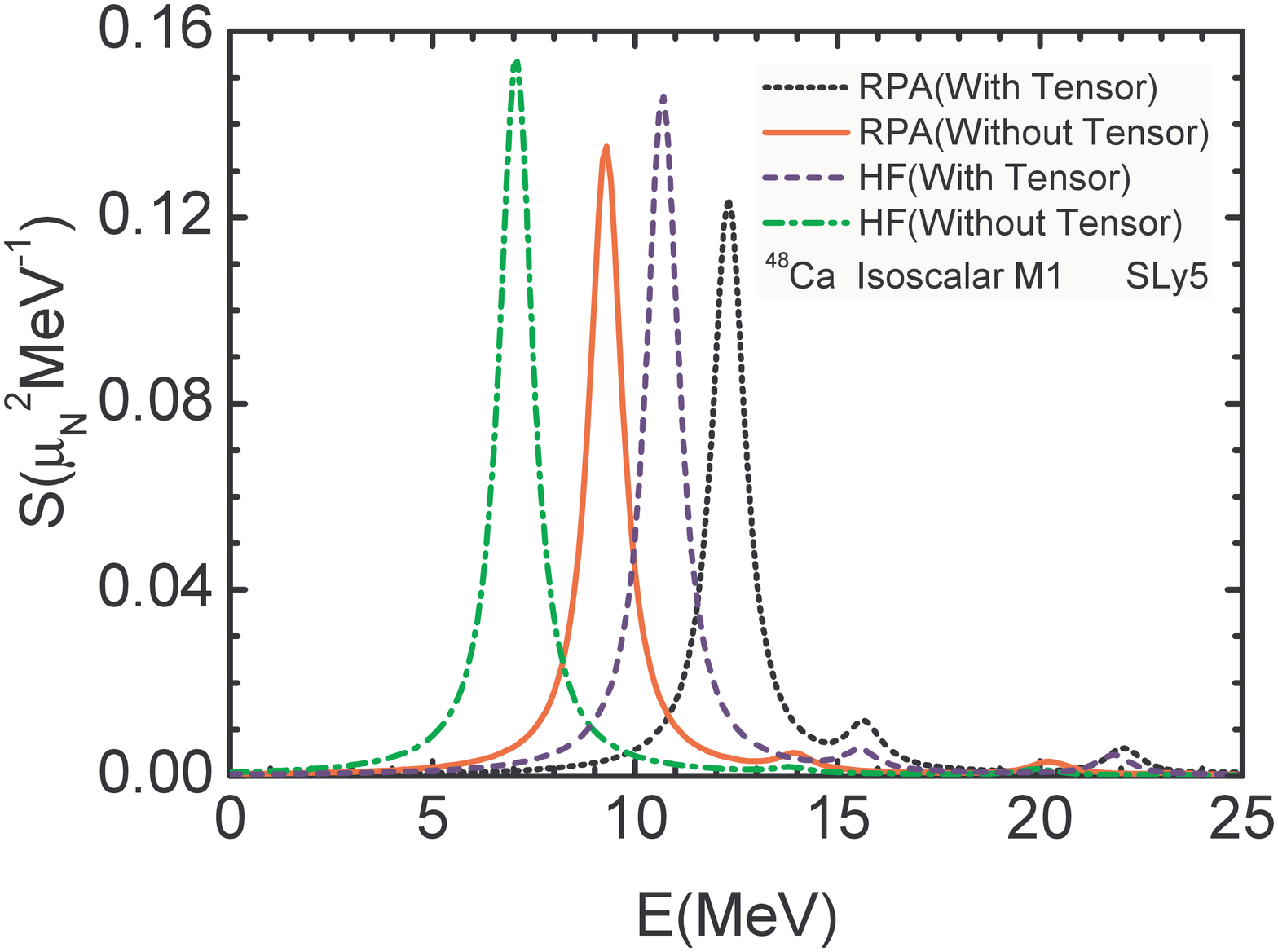}
\includegraphics[width=0.495\textwidth]{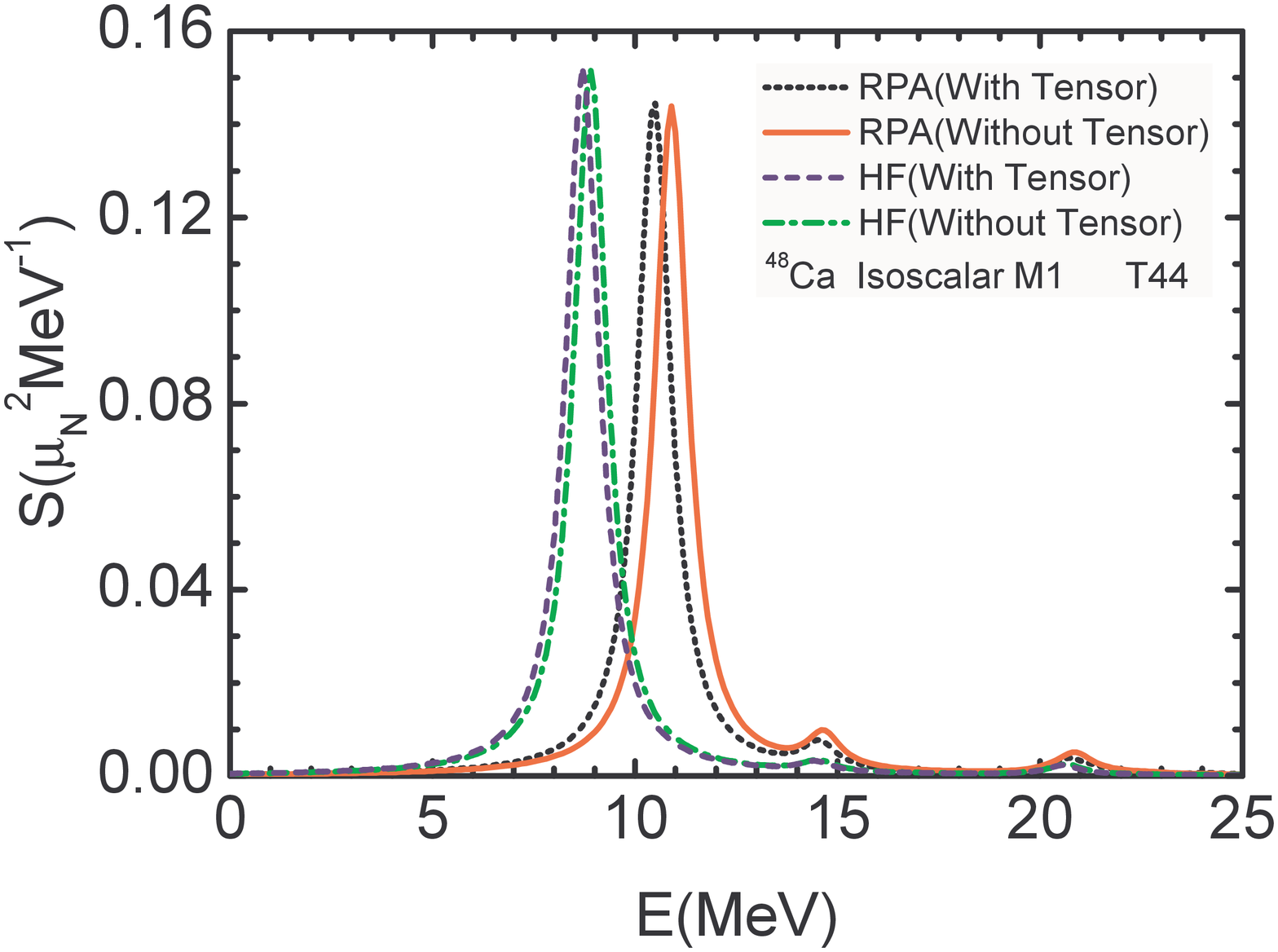}
\vglue -3.0cm
\includegraphics[width=0.495\textwidth]{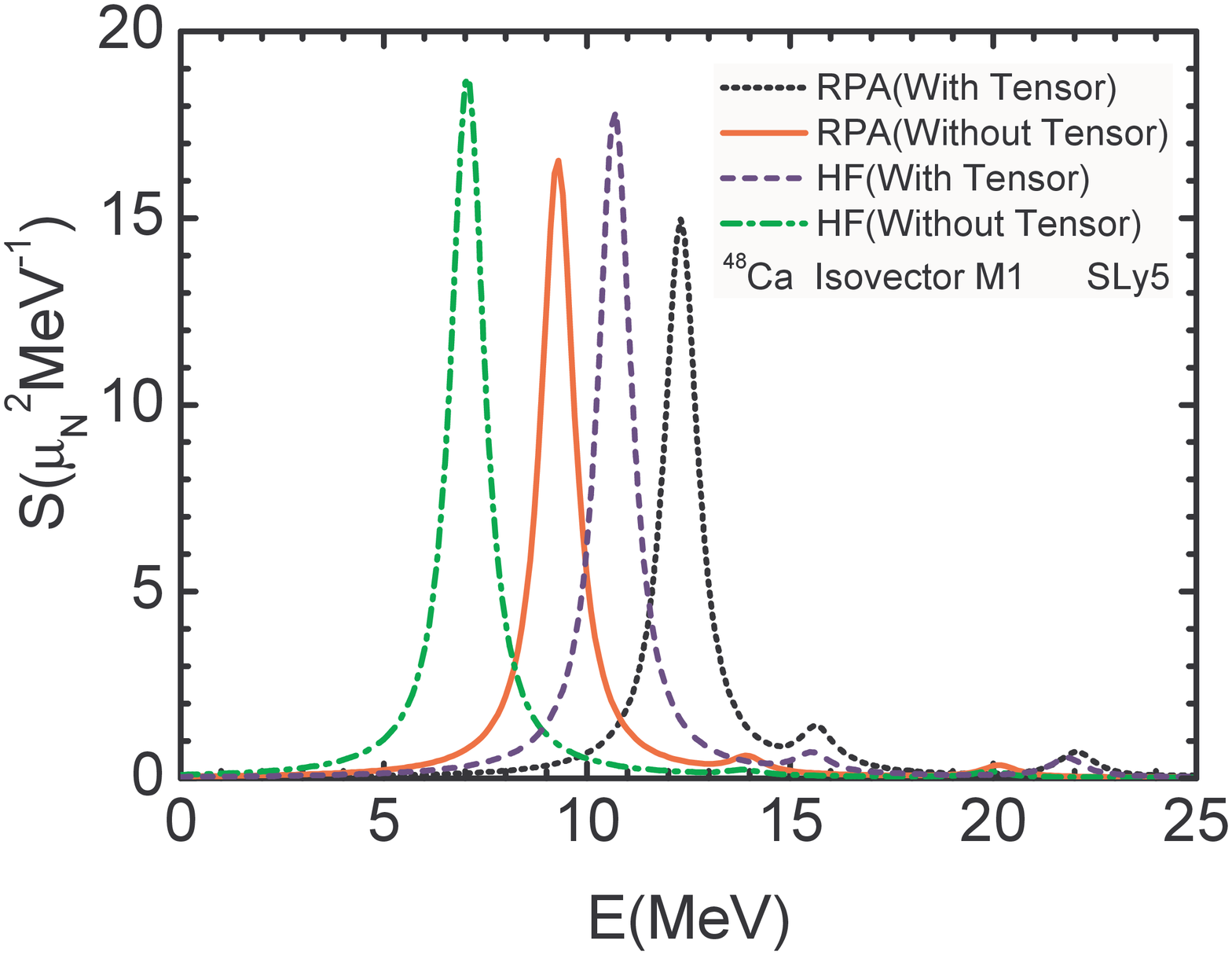}
\includegraphics[width=0.495\textwidth]{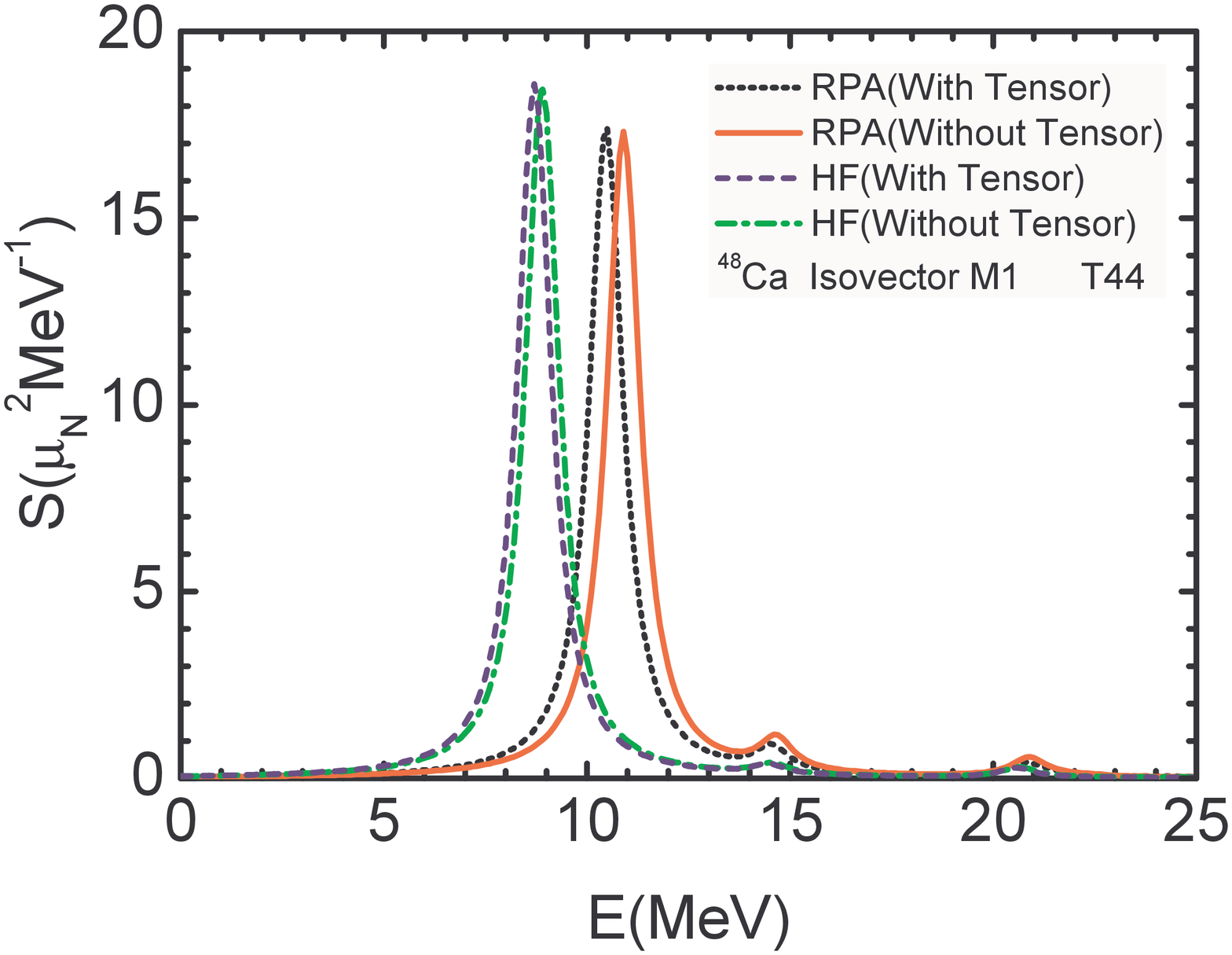}
\vglue -3.0cm \caption{Unperturbed and RPA strength functions 
in $^{48}$Ca associated 
with the isoscalar and isovector M1 operators defined by Eqs. 
(\ref{m1is}) and (\ref{m1iv}). We display results both with and
without the inclusion of the tensor force, in the case of 
SLy5 plus the $T$ and $U$ parameters of \cite{Col.07} 
(left panels) and in the case of
the T44 parameter set (right panels).
The discrete RPA peaks have been 
smeared out by using Lorentzian functions which have 1 MeV width.} 
\label{Fig.6}
\end{figure}

The results we have obtained in the case of $^{48}$Ca are displayed in
Fig. \ref{Fig.6}. In the left panels, the strength functions
associated with the SLy5 Skyrme set plus the $T$ and $U$ 
parameters of \cite{Col.07} are shown. The HF peak is 
associated with the neutron f$_{7/2}\rightarrow$ f$_{5/2}$ 
configuration, which lies at 7.06 MeV without the tensor
contribution, and is pushed at 10.68 MeV by the inclusion of
tensor terms. In the case of RPA, we deal with a somewhat undesired 
feature of the SLy5 set, which is characterized by a positive 
value of the Landau parameter $G_0$=1.14 and a negative value of 
$G_0^\prime$=-0.15. Repulsion in the spin-isoscalar (IS) channel and attraction
in the spin-isovector (IV) channel is in contrast with the empirically
accepted values of the Landau parameters, and, as already 
noticed in \cite{Fra.07}, constitutes an anomaly of the SLy5 set.
In the case at hand, the residual interaction is essentially
the $J^\pi$=1$^+$ diagonal matrix element of the mentioned neutron
configuration, that is, it is the sum of the IS and IV part.
Without tensor, from the fact that the RPA peak is at 9.28 MeV
one can extract that the interaction is dominated by the IS 
contribution and is repulsive (its value being 2.22 MeV). 
Including the tensor interaction, the RPA peak 
moves to 12.31 MeV, so the shift due to tensor correlations
is 3.03 MeV:
the tensor residual interaction can then be extracted from 
Eq. (\ref{DeRPA}) and it turns out to be -0.59 MeV. The fact
that it is attractive is understandable in keeping with 
Eq. (\ref{appb:eq18}) and from the negative sign of $U$.

In the case of the T44 interaction, the values of the
Landau parameters $G_0$ and $G_0^\prime$ are, respectively, 
0.40 and 0.06. The results associated with this
parameter set are visible in the right panels of 
Fig. \ref{Fig.6}, and one can see that the effect of the tensor 
is quite small as far as both unperturbed mean-field and
RPA correlations are concerned. This is related to the smallness of
the parameter $U$ [cf. Table \ref{table_param}, Eq. (\ref{Jq}) and
the formulas of Appendix B]. In the case of the magnetic
dipole state we have, therefore, found that different
ways of including the tensor terms on top of the Skyrme
interaction produce different effects. The experimental
value of the M1 peak is 10.23 MeV \cite{Nan.84}. Since the 
theoretical values are 9.28 MeV (SLy5 without tensor), 
12.31 MeV (SLy5 with tensor), 10.47 MeV (T44), it seems that
the result obtained with T44 is preferable. 

\begin{figure}[hbt]
\includegraphics[width=0.495\textwidth]{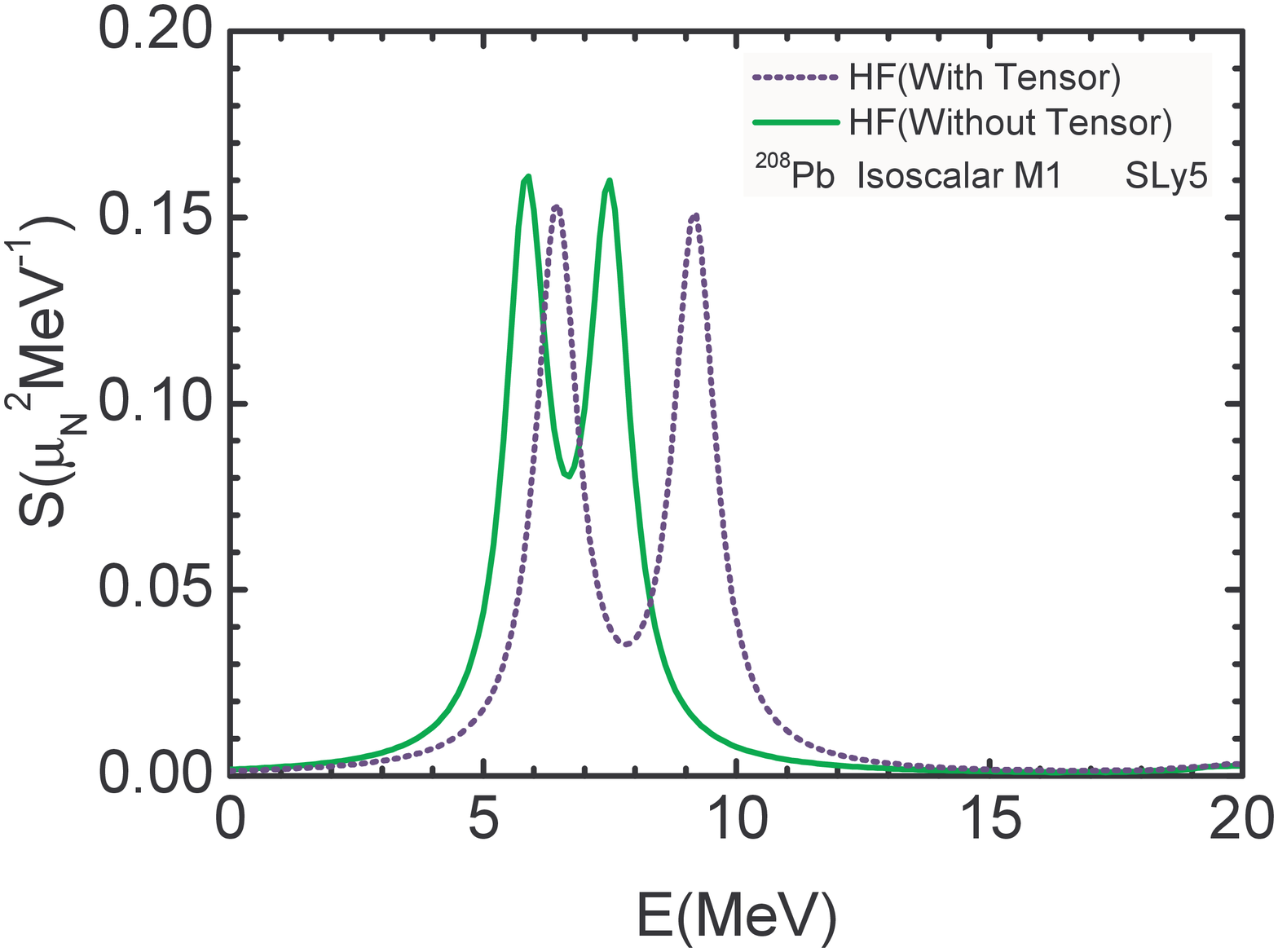}
\vglue -3.0cm
\includegraphics[width=0.495\textwidth]{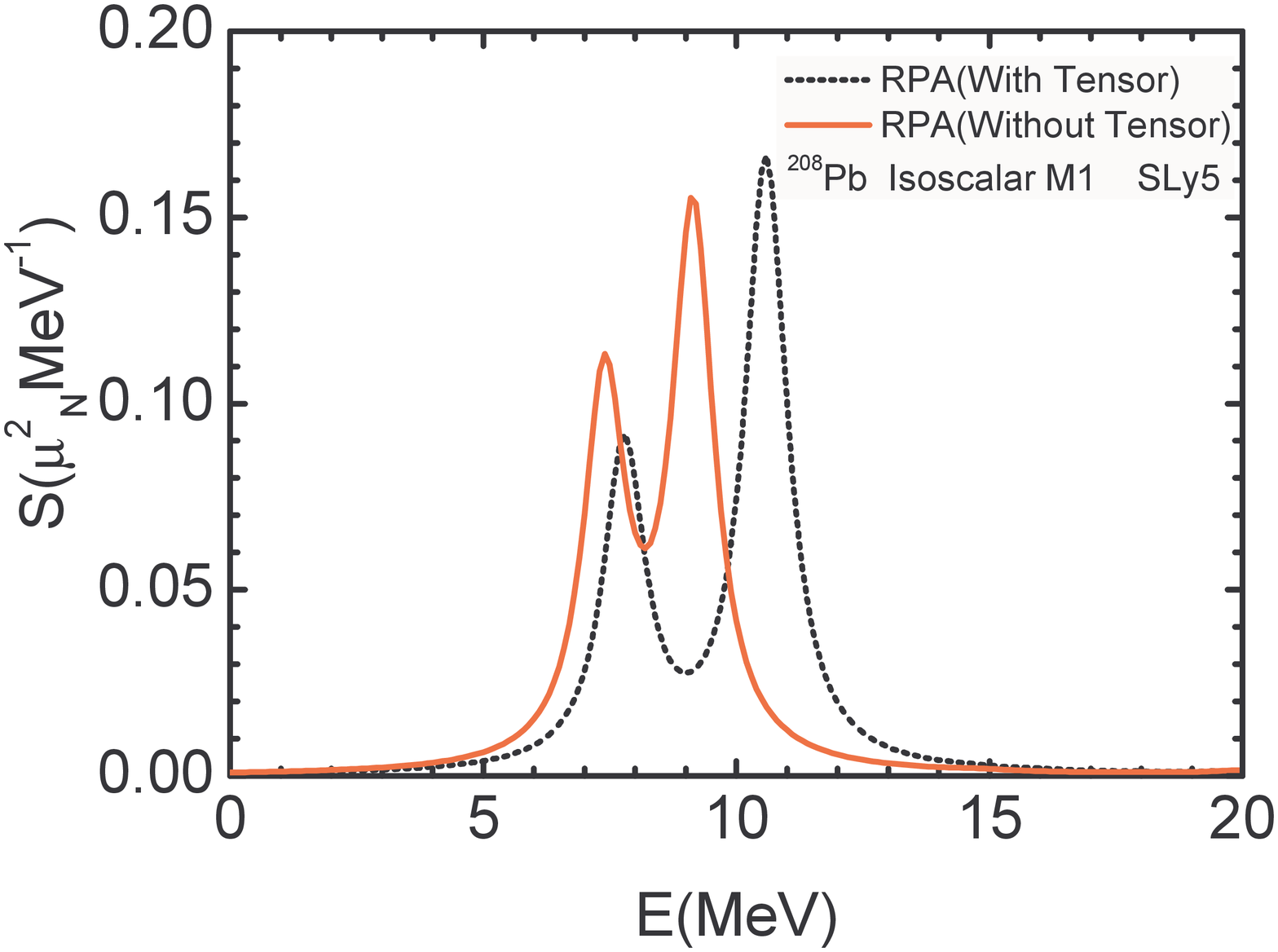}
\vglue -3.0cm
\includegraphics[width=0.495\textwidth]{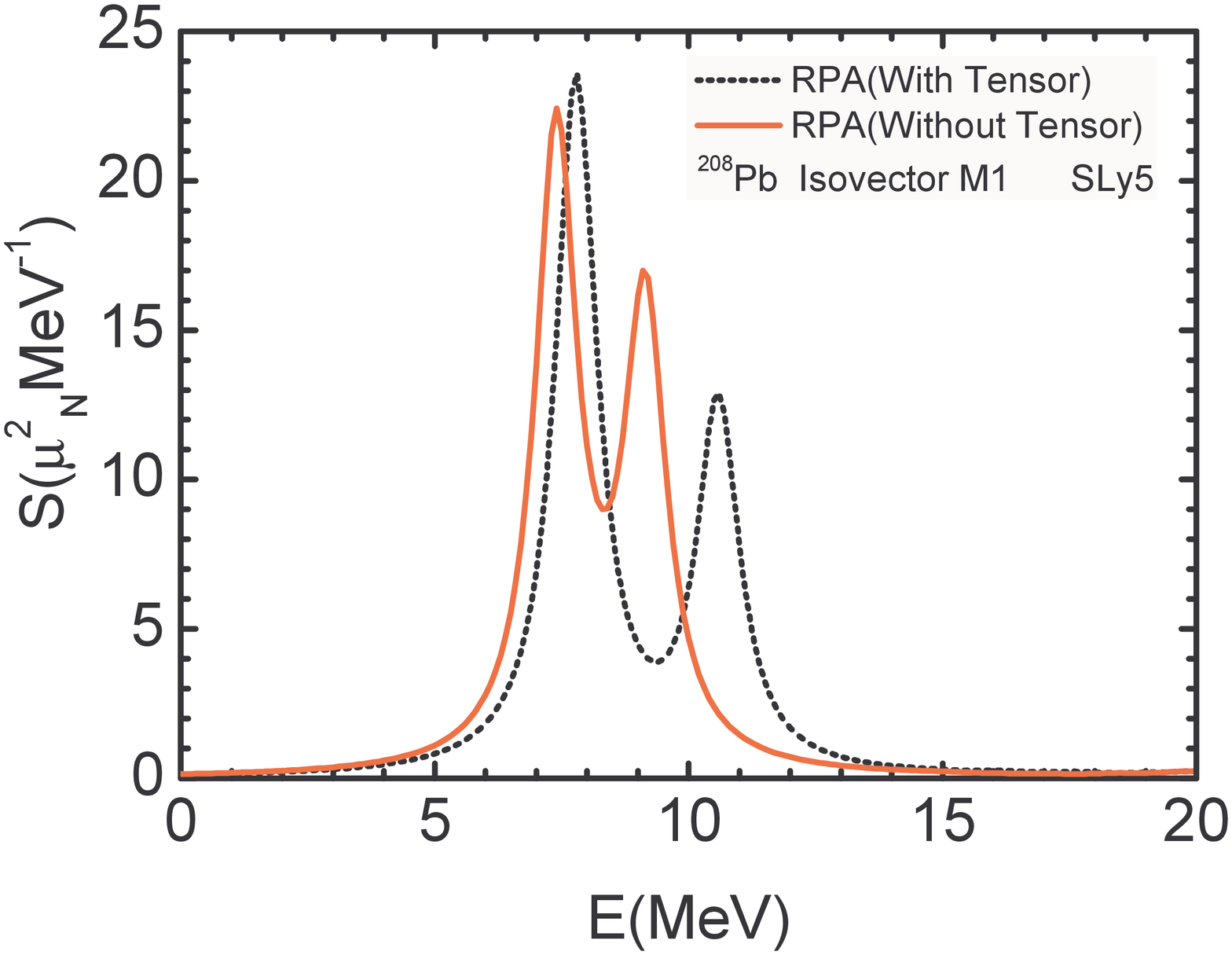}
\vglue -3.0cm \caption{Unperturbed (upper panel) and RPA 
(middle and lower panels) strength functions in $^{208}$Pb 
associated with the isoscalar and isovector M1 operators 
defined by Eqs. (\ref{m1is}) and (\ref{m1iv}). We display results both with and
without the inclusion of the tensor terms, in the case of the
force SLy5 plus the $T$ and $U$ parameters of \cite{Col.07} 
(first column of Table \ref{table_param}). The discrete RPA
peaks have been smeared out by using Lorentzian functions which
have 1 MeV width.} 
\label{Fig.4}
\end{figure}

\begin{figure}[hbt]
\includegraphics[width=0.495\textwidth]{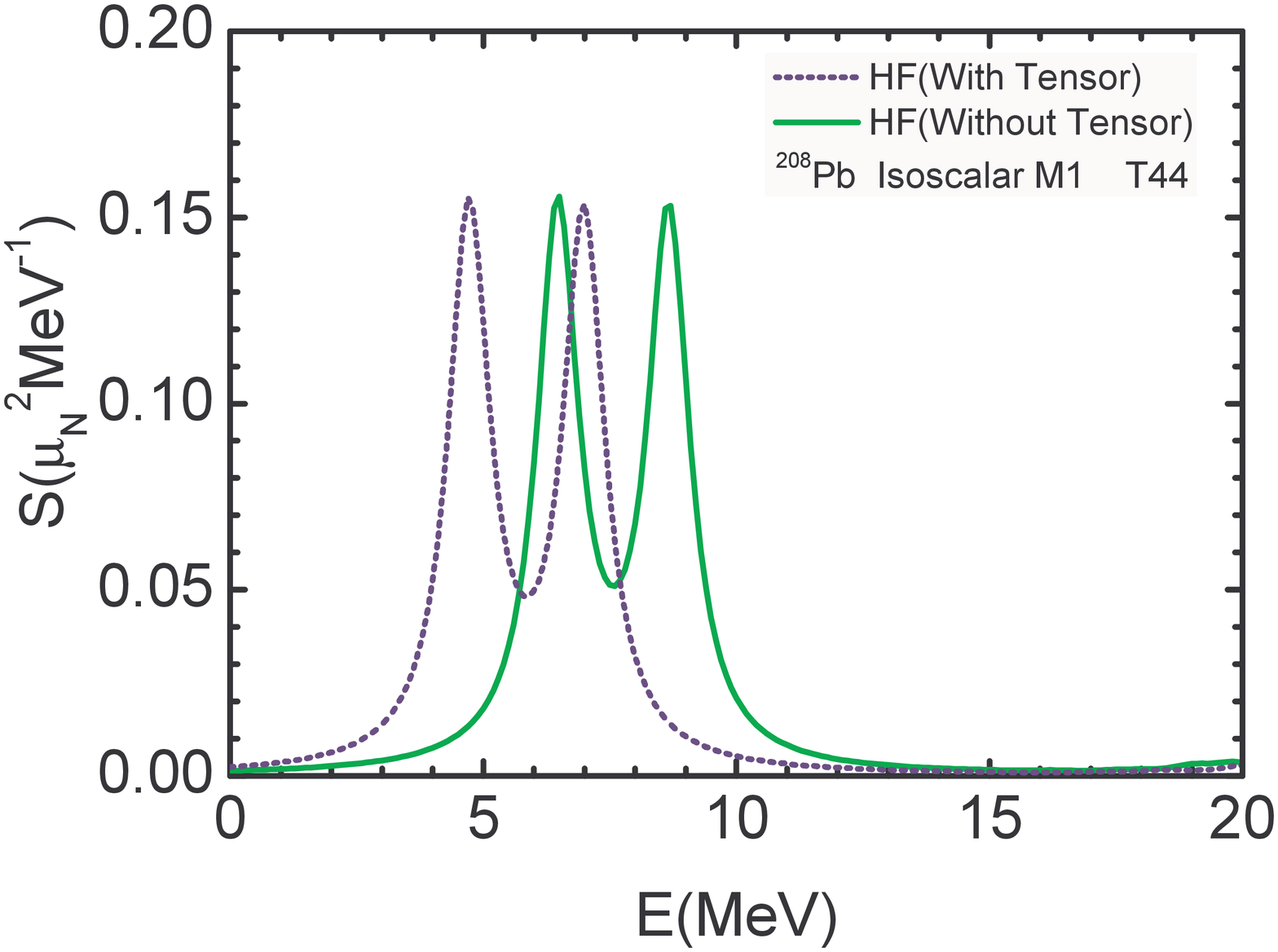}
\vglue -3.0cm
\includegraphics[width=0.495\textwidth]{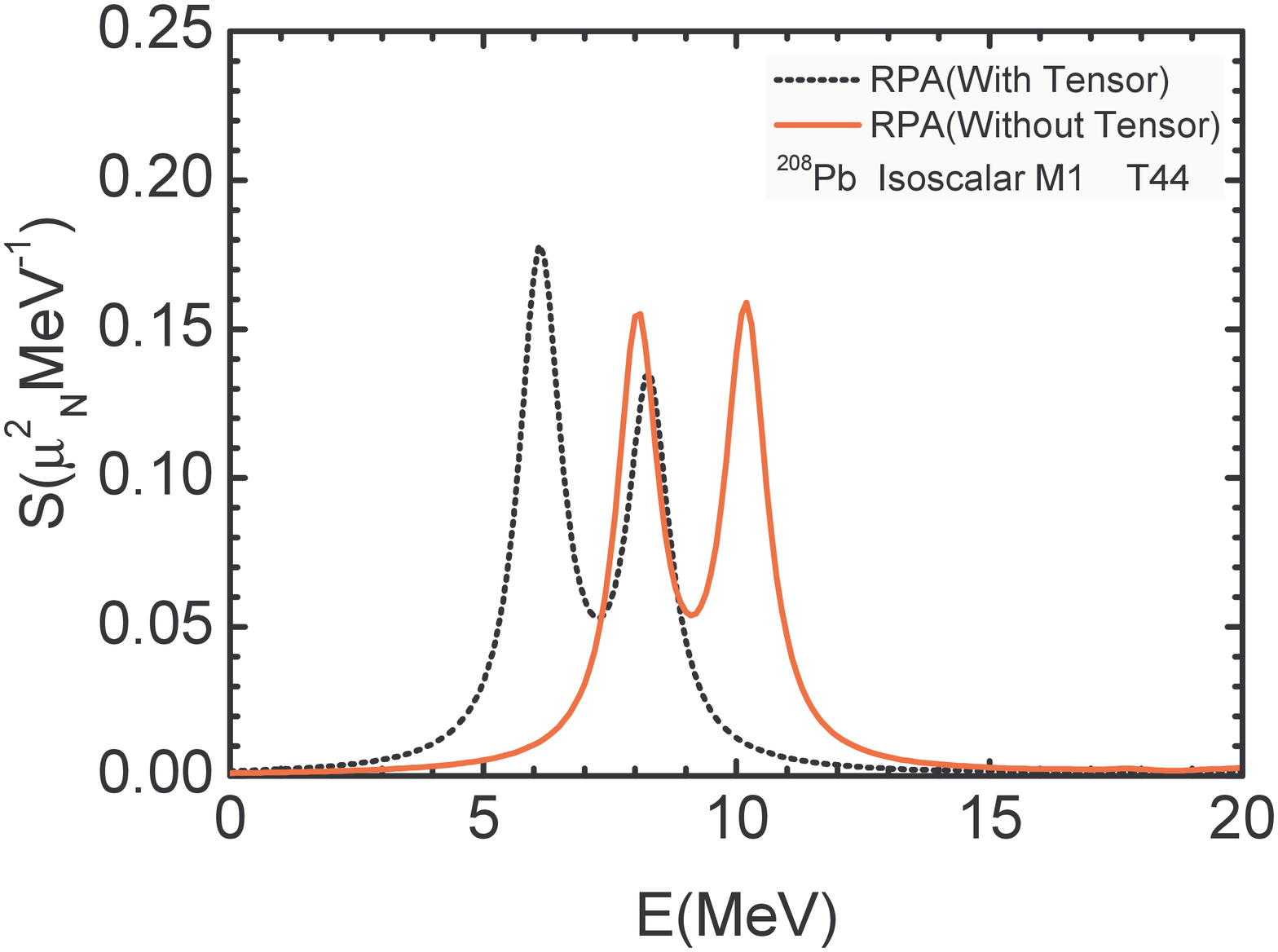}
\vglue -3.0cm
\includegraphics[width=0.495\textwidth]{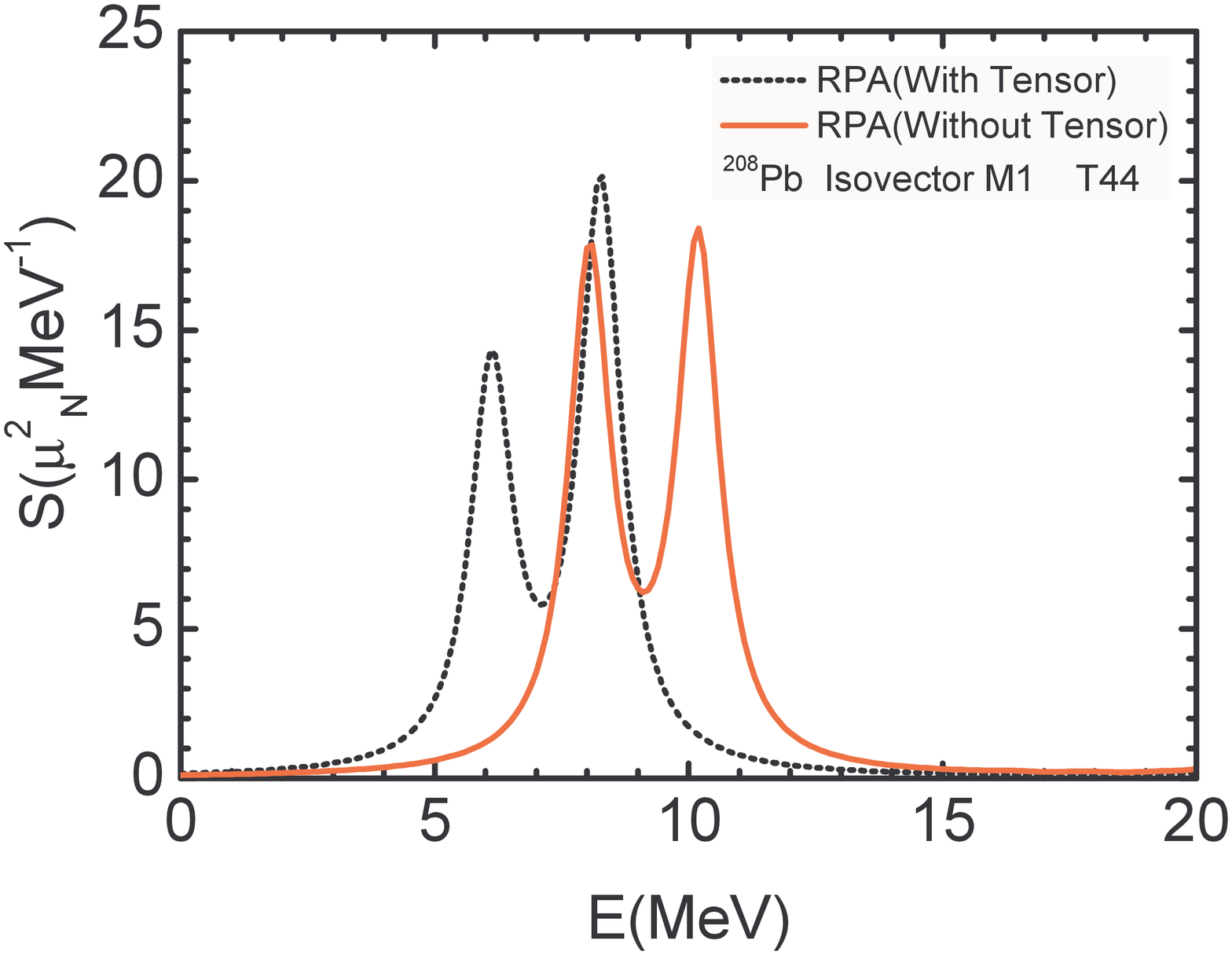}
\vglue -3.0cm \caption{The same as Fig. \ref{Fig.4} in the case of the T44
parameter set (second column of Table \ref{table_param}).} 
\label{Fig.5}
\end{figure}

We turn to the analysis of the results for $^{208}$Pb
which are reported in Figs. \ref{Fig.4} and \ref{Fig.5}, in the 
case of SLy5 plus the tensor force parameters of \cite{Col.07} 
and in the case of T44, respectively. In the
unperturbed spectrum, the main role is played by the proton 
h$_{11/2}\rightarrow$ h$_{9/2}$ and neutron 
i$_{13/2}\rightarrow$ i$_{11/2}$ configurations. They are clearly
visible in the upper panels of the two figures. In the case of the 
SLy5 interaction, without tensor, the proton configuration lies 
at 5.85 MeV while the neutron configuration is at 7.49 MeV. 
As already mentioned several times, in the case of SLy5 plus the 
the tensor parameters of \cite{Col.07} the spin-orbit splittings are
increased when the tensor terms are taken into account: the 
energies become 6.45 MeV and 9.17 MeV, respectively, for the
proton and neutron configurations we have mentioned.  

Within RPA, without tensor, we have 
found two peaks at 7.39 MeV and 9.14 MeV.
The lowest peak is mainly composed by the proton 
h$_{11/2}\rightarrow$ h$_{9/2}$ configuration with an 
admixture of the neutron i$_{13/2}\rightarrow$ i$_{11/2}$ 
configuration having different sign in its amplitude; 
the highest peak is mainly based on the neutron
i$_{13/2}\rightarrow$ i$_{11/2}$ configuration, with some admixture
of the proton one, h$_{11/2}\rightarrow$ h$_{9/2}$, having the same sign in
its amplitude. In other words, the lowest (highest) state has 
more IV (IS) character. This isospin character is not
strongly pronounced because, with the values of $T$ and $U$ that
have been employed, the non-diagonal matrix element which mixes
the proton and neutron configurations is small. 
Since experimentally one finds that the 
lowest (highest) state has more IS (IV) character, there is
a doublet inversion which is related to the 
values of the Landau parameters (discussed above). We also 
notice that inversions of IS and IV spin doublet have also 
emerged from the self-consistent Gogny calculations of Ref. \cite{VDD}. 

With the inclusion of the tensor terms, the lowest and highest
peak move, respectively, to 7.79 MeV and 10.57 MeV. If
we neglect the small mixing between the proton and
neutron configurations, we can extract the values
of the matrix elements of the residual tensor force from 
Eq. (\ref{DeRPA}) separately for the proton and neutron
states. We find, respectively, -0.20 MeV and -0.25 MeV.
These values are smaller but have the same sign as
in $^{48}$Ca. 

Similarly to $^{48}$Ca, the results in $^{208}$Pb 
obtained by using T44 seem better. In particular, 
we can notice that in
this case one has the correct ordering in the spin
doublet, the IS (IV) peak being the lowest (highest).
In fact, with the values of $T$ and $U$ associated with
this parameter set, the IS and IV characters of the
states are more pronounced due to a larger value
of the matrix element mixing neutrons and protons. 
The values of the energies are respectively 6.12 MeV and
8.27 MeV and they compare better with the experimental
findings (see e.g. \cite{Shi.08} and references therein) 
which are 5.85 MeV and 7.30 MeV, than SLy5
without tensor (which gives 7.39 MeV and 9.14 MeV) or
SLy5 with tensor (which gives 7.79 MeV and 10.57 MeV).

\section{Conclusion and perspectives}
\label{conclu}

The tensor component of the bare nuclear force has been always of central
interest for nuclear physics, but as far as effective interactions for
many-nucleon systems are concerned, only recently we have achieved some
first understanding of its role within self-consistent frameworks. 
In particular, in recent years many works have appeared in the
literature which deal with the effect of tensor terms when they are
added on top to, or fitted together with, Skyrme-type forces. These
works have dealt with the impact of tensor terms on bulk properties like
masses, or on single-particle states.

Our work deals, for the first time, with the effects produced by these 
tensor terms on the multipole response of finite nuclei. We have attempted
to disentangle the effects due to the modifications of the static
mean field, and those due to the residual interaction, by analyzing
the results of self-consistent HF plus RPA calculations performed in the
case of quadrupole, octupole and magnetic dipole channels.

The modifications of the static mean field had been already understood
by means of our previous works. The dominant character (attractive or 
repulsive) of the residual interaction matrix elements has been extracted
from the numerical calculations but also understood on the basis of 
a separable approximation for the tensor p-h force. Then, it has become
evident that since these two effects are governed by different 
combinations of the parameters of the tensor force, the effects which
are visible on the final results are ruled by a delicate interplay and
are non-trivial. This is one of the main findings of the present work.

The magnetic dipole states are more affected by the
inclusion of the tensor terms at the mean field level, 
since unperturbed configurations correspond exactly
to energy jumps between spin-orbit partners. The 
inclusion of the tensor residual interaction is
demanded by self-consistency: for its matrix elements
we have extracted typically values around few or
several hundreds of keV, but these values depend of
course on the values of the tensor force parameters
$T$ and $U$. 

Our work is exploratory and we have shown that the tensor
force plays
a role in RPA since the matrix elements are in general
not negligible. We have considered two cases, namely the 
parameter set SLy5 plus the tensor parameters of \cite{Col.07} 
and the set T44. The latter seems preferable if
one looks at the comparison with experiment in the 
magnetic dipole case, but not for 
some of the low-lying states in Ca isotopes. 

Actually, before 
learning more from the comparison with experiment, 
and fitting an ultimate Skyrme set including the tensor
terms, one should ask the question whether the Skyrme
ansatz is general enough for the spin-isospin channel.
Some of the most modern, and most widely used, sets like
the Lyon parameterizations have somewhat unsatisfactory
values of the spin and spin-isospin Landau parameters
and this anomaly should be cured. Improved Skyrme sets
and Skyrme sets fitted with the tensor terms can be
studied in detail, as far as their performance for
excited states is concerned, by using the present
RPA formalism. This is the main perspective opened
by our present work.

\section*{Acknowledgments}

C.L. acknowledges the support of the UniAMO fellowship 
provided by Fondazione Cariplo and 
Universit\`a degli Studi which has allowed his stay in 
Milano, and the support of the National Science Foundation 
of China under Grant Nos. 10875150. C.L. and G.C. also
acknowledge partial support from the Kavli Institute for
Theoretical Physics China (KITPC). 
This work is partially supported by the Japanese  
Ministry of Education, Culture, Sports, Science and Technology 
by Grant-in-Aid for Scientific Research under
the program number (C(2)) 20540277.  

\appendix

\section{Calculation of the p-h matrix elements of the tensor interaction}
\label{appa}

We use the triplet-even and triplet-odd zero-range tensor terms
which have been introduced originally by Skyrme \cite{Skyrme} and read 
\begin{eqnarray}
V_{\rm tensor}=
&&\frac{T}{2}\left\{ [\bf{(\sigma_1 \cdot k^{'})(\sigma_2 \cdot
k^{'})}-\frac{1}{3}(\sigma_1 \cdot
\sigma_2)k^{'2}]\delta(\bf{r_1-r_2)}
+\delta(\bf{r_1-r_2)}[\bf{(\sigma_1 \cdot k)(\sigma_2 \cdot k)}
-\frac{1}{3}(\sigma_1 \cdot \sigma_2)k^{2}]\right\} \nonumber\\
&&+U\left\{ \bf{(\sigma_1 \cdot k^{'})\delta(\bf{r_1-r_2})(\sigma_2
\cdot k)-\frac{1}{3}(\sigma_1 \cdot \sigma_2)\delta(\bf{r_1-r_2})
[k^{'} \cdot k]} \right\},
\label{appa:eq1}
\end{eqnarray}
where
$k^{'}=-\frac{\mathbf{\overleftarrow{\nabla}_1-\overleftarrow{\nabla}_2}}{2i}$
acts on the left while
$k=\frac{\mathbf{\overrightarrow{\nabla}_1-\overrightarrow{\nabla}_2}}{2i}$
acts on the right.
Using the formula
\begin{equation}
\bf{(a \cdot b)(c \cdot d)=\frac{(a \cdot c)(b \cdot
d)}{3}+\frac{(a \times c)(b \times
d)}{2}+\sum^{2}_{q=-2}(-1)^{2-q}[a^{(1)} \otimes
c^{(1)}]_q^{(2)}[b^{(1)} \otimes d^{(1)}]_{-q}^{(2)}}, 
\end{equation}
we can re-write (\ref{appa:eq1}) in such a way that the coupling of both the 
spatial and spin operators to rank 2 is evident,
\begin{eqnarray}
V_{\rm tensor}&=&\frac{T}{2}\left\{ \sum^{2}_{q=-2}(-1)^{2-q}\bf{[\sigma_1 \otimes
\sigma_2]^{(2)}_q[k^{'} \otimes
k^{'}]^{(2)}_{-q}}\delta(\bf{r_1-r_2)} \right. \nonumber\\
&& \left. +\delta(\bf{r_1-r_2)}\sum^{2}_{q=-2}(-1)^{2-q}[\bf{[\sigma_1
\otimes \sigma_2]^{(2)}_q[k \otimes k]^{(2)}_{-q}} \right\} \nonumber\\
&&+U \left\{ \sum^{2}_{q=-2}(-1)^{2-q}\bf{[\sigma_1 \otimes
\sigma_2]^{(2)}_q [k^{'} \otimes k]^{(2)}_{-q}
\delta(\bf{r_1-r_2})} \right\}.
\label{appa:eq3}
\end{eqnarray}
For RPA, we wish to calculate the coupled p-h matrix elements defined as
\begin{equation}
V^{(J)}_{abcd}=
\frac{1}{2J+1}\sum_{m_a,m_b,m_c,m_d,M}(-1)^{j_c-m_c+j_b-m_b}\langle
j_am_a,j_c-m_c|JM\rangle\langle j_dm_d,j_b-m_b|JM\rangle\langle
am_abm_b|V|cm_cdm_d\rangle.
\label{phme}
\end{equation}
We use the well-known Pandya relation to re-express the p-h matrix elements 
in terms of the particle-particle (pp) ones, 
\begin{equation}
V^{(J)}_{abcd}=
\sum_{J^{'}}\widehat{J^{'}}^{2}(-1)^{j_c+j_d+J^{'}} 
\left\{
{{\begin{array}{*{30}c}
J^{'} \hfill & j_b \hfill &  j_a\hfill\\
J \hfill & j_c \hfill & j_d \hfill \\
\end{array} }}\right\}
\langle (j_aj_b)J^{'}|V|(j_cj_d)J^{'} \rangle, 
\end{equation}
and change the jj coupling to the LS one, to obtain 
\begin{eqnarray}
V^{(J)}_{abcd}&=&
\sum_{J^{'}}\widehat{J^{'}}^{2}(-1)^{j_c+j_d+J^{'}} 
\left\{
{{\begin{array}{*{30}c}
J^{'} \hfill & j_b \hfill &  j_a\hfill\\
J \hfill & j_c \hfill & j_d \hfill \\
\end{array} }}\right\}
\nonumber\\
& \times &\sum_{L,L^{'},S,S^{'}} 
\widehat{j_a}\widehat{j_b}\widehat{j_c}\widehat{j_d}\widehat{S}\widehat{S^{'}}
\widehat{L}\widehat{L^{'}}\left\{ {{\begin{array}{*{30}c}
l_a \hfill & L \hfill & l_b \hfill\\
j_a \hfill & J^{'} \hfill & j_b \hfill \\
\frac{1}{2} \hfill & S \hfill & \frac{1}{2} \hfill \\
\end{array} }} \right\}\left\{ {{\begin{array}{*{30}c}
l_c \hfill & L^{'} \hfill & l_d \hfill\\
j_c \hfill & J^{'} \hfill & j_d \hfill \\
\frac{1}{2} \hfill & S^{'} \hfill & \frac{1}{2} \hfill \\
\end{array} }} \right\}\langle
(LS)J^{'}|V|(L^{'}S^{'})J^{'} \rangle.
\end{eqnarray}
Then,
\begin{eqnarray}
\langle (LS)J^{'}M^{'}|V|(L^{'}S^{'})J^{'}M^{'} \rangle
=(-1)^{S+J^{'}+L^{'}}\left\{ {{\begin{array}{*{30}c}
L \hfill & S \hfill & J^{'} \hfill\\
S^{'} \hfill & L^{'} \hfill & 2 \hfill \\
\end{array} }} \right\}\langle S\|\left[ \sigma_1 \otimes \sigma_2 \right]^{(2)} \|S^{'}
\rangle \langle L \|\left[ \hat O_1\otimes \hat O_2 \right]^{(2)} \|L^{'}
\rangle,
\end{eqnarray}
where the operators $\hat O_i$ acts on the space part of the wavefunction.
For the spin part, the reduced matrix element is simply 
\begin{eqnarray}
&&\langle S\|\left[ \sigma_1 \otimes \sigma_2 \right]^{(2)}
\|S^{'} \rangle
=\sqrt{3\times3\times5}\times\frac{1}{9}\times\sqrt{6}\times\sqrt{6}=2\sqrt{5}
\end{eqnarray}
while the space part is quite lengthy to evaluate. After applications of the
gradient formula and proper recouplings, one arrives at the final result, that is,  
\begin{eqnarray}
\langle (LS)J^{'}|V|(L^{'}S^{'})J^{'} \rangle & = & 
\sum_{\alpha,\lambda,\lambda^{'}}(-1)^{S+J^{'}+L^{'}}\left\{
{{\begin{array}{*{30}c}
L \hfill & 1 \hfill & J^{'} \hfill\\
1 \hfill & L^{'} \hfill & 2 \hfill\\
\end{array} }} \right\}
\times2\sqrt{5}
\nonumber\\
& \times &
\left\{ {{\begin{array}{*{30}c}
l_a \hfill & l_b \hfill & L \hfill\\
l_c \hfill & l_d \hfill & L^{'} \hfill \\
\lambda' \hfill & \alpha \hfill & 2 \hfill \\
\end{array} }} \right\} \widehat{\lambda'}^2\widehat{\lambda}^2
\times\widehat{L}\widehat{L^{'}}\times 5
\times\frac{1}{4} \sum_{n=1}^{10} v^{(n)},
\end{eqnarray}
where
\begin{eqnarray}
v^{(1)}
&=& \frac{T}{2} \sum_{k,i=\pm 1}
(-1)^{\frac{i}{2}+\frac{k}{2}+k+\lambda+\lambda'}
\left(l_a+\frac{i}{2}+\frac{1}{2}\right)^{\frac{1}{2}}
\left(l_a+i+\frac{k}{2}+\frac{1}{2}\right)^{\frac{1}{2}} \left\{
{{\begin{array}{*{30}c}
1 \hfill & 1 \hfill & 2 \hfill\\
\lambda' \hfill & \alpha \hfill & \lambda \hfill \\
\end{array} }} \right\} \left\{ {{\begin{array}{*{30}c}
l_c\hfill & l_a+i \hfill & \lambda \hfill\\
1 \hfill & \alpha  \hfill & l_a+i+k \hfill \\
\end{array} }} \right\}
\nonumber \\
& \times & \left\{ {{\begin{array}{*{30}c}
1 \hfill & \lambda \hfill & \lambda'\\
l_c \hfill & l_a \hfill & l_a+i \hfill \\
\end{array} }} \right\} \langle
l_c\|Y_\alpha\|l_a+i+k\rangle\langle l_b\| Y_{\alpha}\|l_d\rangle
\int dr r^2 \phi_b \phi_c(r)\phi_d(r) D_{l_a}^i
D_{l_a+i}^k\phi_a(r),
\label{appa:v1}
\end{eqnarray}
$v^{(2)}$ has the same form as $v^{(1)}$ with the changes 
$a \longleftrightarrow b$ and $c \longleftrightarrow d$, 
$v^{(3)}$ has the same form as $v^{(1)}$ with the change
$a \longleftrightarrow c$ and an additional phase 
$(-1)^{\lambda'}$, and 
$v^{(4)}$ has the same form as $v^{(1)}$ with the changes 
$a \longleftrightarrow d$ and $c \longleftrightarrow b$ plus
the additional phase $(-1)^{\lambda'-\alpha}$. Then
\begin{eqnarray}
v^{(5)}
&=& U \sum_{k,i=\pm 1}
(-1)^{\frac{i}{2}+\frac{k}{2}-i+1+\alpha }
\left(l_a+\frac{i}{2}+\frac{1}{2}\right)^{\frac{1}{2}}
\left(l_c+\frac{k}{2}+\frac{1}{2}\right)^{\frac{1}{2}}
\left\{ {{\begin{array}{*{30}c}
1 \hfill & 1 \hfill & 2 \hfill\\
\lambda' \hfill & \alpha \hfill & \lambda \hfill \\
\end{array} }} \right\}
\left\{ {{\begin{array}{*{30}c}
l_c+k\hfill & l_a  \hfill & \lambda \hfill\\
1 \hfill & \alpha  \hfill & l_a+i \hfill \\
\end{array} }} \right\}
\hfill\nonumber\\
& \times & \left\{ {{\begin{array}{*{30}c}
\lambda \hfill & 1 \hfill & \lambda' \hfill\\
l_c \hfill & l_a \hfill & l_c+k \hfill \\
\end{array} }} \right\} 
\langle l_a+i\|Y_{\alpha}\|l_c+k\rangle \langle l_b\|
Y_{\alpha}\|l_d\rangle  \int dr r^2 \phi_b(r)\phi_d(r)
D_{l_a}^i\phi_a(r)
D_{l_c}^k \phi_c(r),
\label{appa:v5}
\end{eqnarray}
and $v^{(6)}$ has the same form as $v^{(5)}$ with the changes 
$a \longleftrightarrow b$ and $c \longleftrightarrow d$.
Finally,  
\begin{eqnarray}
v^{(7)}
&=& T \sum_{k,i=\pm 1}
(-1)^{\frac{i}{2}+\frac{k}{2}+\lambda^{'}}
\left(l_c+\frac{i}{2}+\frac{1}{2}\right)^{\frac{1}{2}}
\left(l_d+\frac{k}{2}+\frac{1}{2}\right)^{\frac{1}{2}}
\left\{ {{\begin{array}{*{30}c}
1 \hfill &\lambda \hfill &\alpha \hfill\\
\lambda^{'}  \hfill & 1 \hfill & 2 \hfill \\
\end{array} }} \right\}  \left\{ {{\begin{array}{*{30}c}
\lambda\hfill & l_a \hfill & l_c \hfill\\
l_c+i  \hfill & 1 \hfill & \alpha \hfill \\
\end{array} }} \right\}  
\hfill\nonumber \\
& \times &
\left\{
{{\begin{array}{*{30}c}
\lambda^{'}\hfill & l_b \hfill & l_d \hfill\\
l_d+k  \hfill & 1 \hfill & \alpha \hfill \\
\end{array} }} \right\} 
\langle l_a\|Y_\alpha\|l_c+i\rangle \langle
l_b\|Y_\alpha\|l_d+k\rangle \int dr r^2 \phi_a(r)
\phi_b(r)D_{l_c}^i\phi_c(r) D_{l_d}^k
\phi_d(r),
\label{appa:v7}
\end{eqnarray}
and $v^{(8)}$ has the same form as $v^{(7)}$ with the changes 
$c \longleftrightarrow a$, $d \longleftrightarrow b$ plus
the additional phase $(-1)^{\lambda-\lambda'}$, 
$v^{(9)}$ has the same form as $v^{(7)}$ with the changes
$T \longrightarrow U$, $c \longleftrightarrow a$ plus
the additional phase $(-1)^{1+\lambda}$, and 
$v^{(10)}$ has the same form as $v^{(7)}$ with the changes 
$T \longrightarrow U$, $b \longleftrightarrow d$ plus the
additional phase $(-1)^{1-\lambda'}$. 

In all the formulas (\ref{appa:v1}), (\ref{appa:v5}) and 
(\ref{appa:v7}) the following differential operator is used:
\begin{equation}
D^m_l = \frac{d}{dr} + \frac{(-)^{\frac{m}{2}+\frac{l}{2}}(l-
\frac{m}{2}+\frac{1}{2})}{r}.
\end{equation}

\section{A separable approximation for the tensor interaction}
\label{appb}

The tensor-even term in Eq. (\ref{appa:eq3}) can be written
as 
\begin{equation}
V^{\rm even}_{\rm tensor}=\frac{T}{2} \{ [
\sqrt{5} [{\bf\sigma}_1 \otimes {\bf\sigma}_2]^{(2)} 
\otimes 
[{\bf k}^{'} \otimes {\bf k}^{'}]^{(2)} ]^{(0)} + \sqrt{5} [
[{\bf\sigma}_1 \otimes {\bf\sigma}_2]^{(2)} 
\otimes 
[{\bf k} \otimes {\bf k}]^{(2)} ]^{(0)}
\} \delta(\bf{r_1-r_2)}. 
\end{equation}
We assume that the contributions including gradients acting
twice on the same wavefunction ($\sim{\bf\nabla}_1\cdot
{\bf\nabla}_1$
or $\sim{\bf\nabla}_2\cdot{\bf\nabla}_2$) are less important
than the terms $\sim{\bf\nabla}_1\cdot{\bf\nabla}_2$ or
$\sim{\bf\nabla}_2\cdot{\bf\nabla}_1$, and we neglect them.
With this approximation, and the multipole expansion of
the $\delta$-function, we can write one of the leading terms
as
\begin{equation}
V^{\rm even}_{\rm tensor}({\rm term\ 1}) = 
\frac{T}{2}
\sum_\ell \sqrt{5} \hat\ell (-)^\ell 
\frac{1}{4} \{ [ {\bf\sigma}_1 \otimes {\bf\sigma}_2 ]^{(2)} 
\otimes [ {\bf\nabla}_1 \otimes {\bf\nabla_2} ]^{(2)} 
\}^{(0)} [ Y_\ell(1) \otimes Y_\ell(2) ]^{(0)}. 
\label{appb:eq2}
\end{equation}
We apply standard angular momentum techniques to express
this term of the interaction as a sum of separable terms,
that is, terms made up with a product of two operators
acting respectively only on particle 1 and particle 2.
The result is
\begin{equation}
V^{\rm even}_{\rm tensor}({\rm term\ 1}) = 
\frac{5T}{8} \sum_{\ell,\lambda,\lambda',k} (-)^{\lambda
+\lambda'+k} \hat\lambda\hat\lambda'\hat k 
\left\{ \begin{array}{ccc} \lambda & \lambda' & 2 \\
1 & 1 & \ell \end{array} \right\} 
\left\{ \begin{array}{ccc} 1 & 1 & 2 \\
\lambda' & \lambda & k \end{array} \right\} 
[ \hat O_{\lambda',k}(1) \otimes \hat O_{\lambda,k}(2) 
]^{(0)},
\label{appb:eq3}
\end{equation}
where the operator is
\begin{equation}
\hat O_{\lambda,k}(i) = [ \sigma_i \otimes ( 
{\bf \nabla}_i \otimes Y_\ell(i) )^{(\lambda)} 
]^{(k)}.
\label{appb:eq4}
\end{equation}
If we calculate the p-h matrix elements coupled to $J$ 
[cf. the previous appendix, Eq. (\ref{phme})], 
then $k=J$ and 
\begin{equation}
V^{(J)}_{p'hh'p}({\rm term\ 1}) =
\frac{5T}{8} 
\sum_{\ell,\lambda,\lambda'} (-)^{\lambda+\lambda'+k} 
\hat\lambda\hat\lambda'\hat k 
\left\{ \begin{array}{ccc} \lambda & \lambda' & 2 \\
1 & 1 & \ell \end{array} \right\} 
\left\{ \begin{array}{ccc} 1 & 1 & 2 \\
\lambda' & \lambda & k \end{array} \right\} 
\langle p' \vert \hat O_{\lambda',k} \vert h' \rangle
\langle h  \vert \hat O_{\lambda,k}  \vert p  \rangle.
\label{appb:eq5}
\end{equation}
We then analyze the quantity
\begin{equation}
\langle p' \vert \hat O_{\lambda',k} \vert h' \rangle
\langle h  \vert \hat O_{\lambda,k}  \vert p  \rangle.
\end{equation}
Due to the time-reversal properties of the operator
(\ref{appb:eq4}), this is equal to
\begin{equation}
- \langle p' \vert \hat O_{\lambda',k} \vert h' \rangle
  \langle p  \vert \hat O_{\lambda,k}  \vert h  \rangle
\label{appb:eq6}
\end{equation}
(cf., e.g., p. 312-313 of \cite{BM1}). To make an 
estimate of the sign of the diagonal
matrix elements, we obtain from (\ref{appb:eq5}) and 
(\ref{appb:eq6})
\begin{equation}
V^{(J)}_{phhp}({\rm term\ 1}) =
- \frac{5T}{8} 
\sum_{\ell,\lambda,\lambda'} (-)^{\lambda+\lambda'+k} 
\hat\lambda\hat\lambda'\hat k 
\left\{ \begin{array}{ccc} \lambda & \lambda' & 2 \\
1 & 1 & \ell \end{array} \right\} 
\left\{ \begin{array}{ccc} 1 & 1 & 2 \\
\lambda' & \lambda & k \end{array} \right\} 
\langle p  \vert \hat O_{\lambda',k} \vert h \rangle
\langle p  \vert \hat O_{\lambda,k}  \vert h  \rangle.
\end{equation}

We apply this latter equation for two typical cases (of interest
for the low-lying spectroscopy of nuclei), namely the
natural parity case $J^\pi$=2$^+$ and the unnatural parity
case $J^\pi$=1$^+$. We obtain
\begin{eqnarray}
{\rm sign} \left(V^{(2^+)}_{phhp}({\rm even, term\ 1})\right) & = &
- {\rm sign} \left( T \right), \label{appb:eq9} \\
{\rm sign} \left(V^{(1^+)}_{phhp}({\rm even, term\ 1})\right) & = &
+ {\rm sign} \left( T \right). \label{appb:eq10}
\end{eqnarray}

Exactly the same kind of study can be performed for the tensor-odd 
part of the interaction, namely for the term
\begin{equation}
V^{\rm odd}_{\rm tensor} = U \{ \sqrt{5} [
[{\bf\sigma}_1 \otimes {\bf\sigma}_2]^{(2)} 
\otimes 
[{\bf k}^{'} \otimes {\bf k}]^{(2)} ]^{(0)}
\} \delta(\bf{r_1-r_2)}. 
\end{equation}
In this case, we assume that the leading term is $\sim \nabla^\prime_1
\cdot \nabla_2$ (where the prime means that the gradient is
acting at left). The equation corresponding to (\ref{appb:eq2}) is
\begin{equation}
V^{\rm odd}_{\rm tensor}({\rm term\ 1}) = 
-\frac{U}{4}
\sum_\ell \sqrt{5} \hat\ell (-)^\ell 
\{ [ {\bf\sigma}_1 \otimes {\bf\sigma}_2 ]^{(2)} 
\otimes [ {\bf\nabla}^\prime_1 \otimes {\bf\nabla_2} ]^{(2)} 
\}^{(0)} [ Y_\ell(1) \otimes Y_\ell(2) ]^{(0)}. 
\end{equation}
By developing the same procedure outlined for the even term, we come
to the conclusion that 
\begin{eqnarray}
{\rm sign} \left(V^{(2^+)}_{phhp}({\rm odd, term\ 1})\right) & = &
+ {\rm sign} \left( U \right), \label{appb:eq13} \\
{\rm sign} \left(V^{(1^+)}_{phhp}({\rm odd, term\ 1})\right) & = &
+ {\rm sign} \left( U \right). \label{appb:eq14}
\end{eqnarray}

We can use Eqs. (\ref{appb:eq9}), (\ref{appb:eq10}), (\ref{appb:eq13}) 
and (\ref{appb:eq14}) as a useful guideline. In fact, these equations
correspond to direct (i.e., non antisymmetrized) matrix elements.
The antisymmetrized interaction can be written as
\begin{equation}
V_{\rm tensor}\left( 1 - P_M P_\sigma P_\tau \right), 
\end{equation}
where the position, spin and isospin exchange operators $P_M$, $P_\sigma$
and $P_\tau$ have been introduced. Since the tensor interaction acts
only among spin-triplet states, $P_\sigma$=1. By definition, $P_M$=+1
(-1) in the case of the even (odd) tensor term. Therefore, if we
call for simplicity 
\begin{eqnarray}
V({\rm even}) & = & V^{(J^\pi)}_{phhp}({\rm even}), \\
V({\rm odd}) & = & V^{(J^\pi)}_{phhp}({\rm odd}), 
\end{eqnarray}
it is simple to arrive at
\begin{eqnarray}
V^{(J^\pi)}_{phhp} & = & \left( \frac{1}{2}V({\rm even}) 
+ \frac{3}{2}V({\rm odd}) \right) 
\ \ \ \ \ \ \ \ \ \ \ \ \ \ \ \ \ \ \ \ {\rm IS\ part} \nonumber \\
& + & \left( -\frac{1}{2}V({\rm even}) 
+ \frac{1}{2}V({\rm odd}) \right) \langle \tau_1\cdot\tau_2 \rangle 
\ \ \ \ \ \ \ \ {\rm IV\ part}. \label{appb:eq18}
\end{eqnarray}
The sign of the residual interaction can be understood by combining
this latter equation with the previous ones. This is at least true when
the different terms add up coherently; in case of cancellations, of course,
the qualitative arguments of this Appendix may not be valid.

\newpage

\begin{table}
\caption{Parameters of the effective interactions used in the HF plus
RPA calculations. The set SLy5 has been introduced in \cite{Cha.98}, and 
it is here supplemented by the tensor force parameters $T$ and $U$ which 
have been introduced in \cite{Col.07}. The set T44 is taken from \cite{Les.07}. 
The notation is the same as in \cite{Cha.98} for the Skyrme parameters, 
while $T$ and $U$ are defined as in \cite{Col.07}.}
\begin{ruledtabular}
\begin{tabular}{ccc}
                   & SLy5+(T,U)      & T44                  \\
\hline
$t_0$ [MeV fm$^3$] & -2484.88        & -2485.670            \\
$t_1$ [MeV fm$^5$] &   483.13        &   494.477            \\
$t_2$ [MeV fm$^5$] &  -549.40        &  -337.961            \\
$t_3$ [MeV fm$^{3+3\sigma}$] 
                   & 13763.0         & 13794.7              \\
$x_0$              &     0.778       &     0.721557         \\
$x_1$              &    -0.328       &    -0.661848         \\
$x_2$              &    -1.000       &    -0.803184         \\
$x_3$              &     1.267       &     1.175908         \\
$\sigma$           &     1/6         &     1/6              \\
$W_0$ [MeV fm$^5$] &   126.0         &   161.367            \\
\hline
$T$   [MeV fm$^5$] &   888.0         &   520.983            \\
$U$   [MeV fm$^5$] &  -408.0         &    21.522            \\
\end{tabular}
\end{ruledtabular}
\label{table_param}
\end{table}

\newpage

\begin{table}
\caption{Properties of the low-lying quadrupole states in
$^{48}$Ca and $^{208}$Pb. Energies and B(E2) values from
the RPA calculations are compared with experimental data
from Ref. \cite{Ram.01}.}
\begin{ruledtabular}
\begin{tabular}{ccc}
                   & Energy [MeV]    & B(E2) [e$^2$fm$^4$]  \\
\hline
$^{48}$Ca          &                 &                      \\
\hline
SLy5 (no tensor)   &  3.05           &     56.5             \\
SLy5 (with tensor) &  3.88           &     52.9             \\
T44 (no tensor)    &  3.35           &     53.2             \\
T44 (with tensor)  &  3.17           &     51.8             \\
Exp.               &  3.83           &     95.0             \\
\hline
$^{208}$Pb         &                 &                      \\
\hline
SLy5 (no tensor)   &  4.89           &     2.94 10$^3$      \\
SLy5 (with tensor) &  5.06           &     3.07 10$^3$      \\
T44 (no tensor)    &  5.04           &     2.86 10$^3$      \\
T44 (with tensor)  &  4.49           &     2.70 10$^3$      \\
Exp.               &  4.07           &     2.97 10$^3$      \\
\end{tabular}
\end{ruledtabular}
\label{table_2plus}
\end{table}

\newpage

\begin{table}
\caption{The same as Table \ref{table_2plus} for the case of
the low-lying octupole states. In this case the nuclei are $^{40}$Ca, 
$^{48}$Ca and $^{208}$Pb. The experimental data are from Ref. 
\cite{Kib.02}.}
\begin{ruledtabular}
\begin{tabular}{ccc}
                   & Energy [MeV]    & B(E3) [e$^2$fm$^6$]  \\
\hline
$^{40}$Ca          &                 &                      \\
SLy5 (no tensor)   &  3.78           &     1.32 10$^4$      \\
SLy5 (with tensor) &  3.02           &     0.91 10$^4$      \\
T44 (no tensor)    &  1.36           &     1.23 10$^4$      \\
T44 (with tensor)  &  1.22           &     1.03 10$^4$      \\
Exp.               &  3.74           &     2.04 10$^4$      \\
\hline
$^{48}$Ca          &                 &                      \\
SLy5 (no tensor)   &  4.78           &     0.54 10$^4$      \\
SLy5 (with tensor) &  6.16           &     0.66 10$^4$      \\
T44 (no tensor)    &  3.44           &     0.42 10$^4$      \\
T44 (with tensor)  &  4.93           &     0.34 10$^4$      \\
Exp.               &  4.51           &     1.00 10$^4$      \\
\hline
$^{208}$Pb         &                 &                      \\
SLy5 (no tensor)   &  3.51           &     6.90 10$^5$      \\
SLy5 (with tensor) &  3.49           &     6.16 10$^5$      \\
T44 (no tensor)    &  3.49           &     5.87 10$^5$      \\
T44 (with tensor)  &  3.19           &     6.29 10$^5$      \\
Exp.               &  2.61           &(5.30$\pm$0.30)10$^5$ \\
\end{tabular}
\end{ruledtabular}
\label{table_3minus}
\end{table}

\newpage

\newpage

\newpage

\newpage

\newpage

\end{document}